\documentclass[iop]{emulateapj-rtx4}
\usepackage{bm}
\usepackage{threeparttable}
\usepackage{CJK}

\def\mbh{M_{\rm{BH}}}
\def\msun{M_{\odot}}

\begin{document}
\begin{CJK*}{UTF8}{gbsn}


\author{Yan-Fei Jiang (姜燕飞)\altaffilmark{1}, 
Paul J. Green\altaffilmark{2}, 
Jenny E. Greene\altaffilmark{3},
Eric Morganson\altaffilmark{4}, 
Yue Shen\altaffilmark{4,5,16}, 
Anna Pancoast\altaffilmark{2,15}, 
Chelsea L. MacLeod\altaffilmark{2},
Scott~F.~Anderson\altaffilmark{6}, 
W. N. Brandt\altaffilmark{7,8,14},
C. J. Grier\altaffilmark{7,8},
H.-W. Rix\altaffilmark{9},
John~J.~Ruan\altaffilmark{6},
Pavlos~Protopapas\altaffilmark{10},
Caroline~Scott\altaffilmark{11},
W. S. Burgett\altaffilmark{12},
K. W. Hodapp\altaffilmark{12},
M. E. Huber\altaffilmark{12},
N. Kaiser\altaffilmark{12},
R. P. Kudritzki\altaffilmark{12},
E. A. Magnier\altaffilmark{12},
N. Metcalfe\altaffilmark{13},
J. T. Tonry\altaffilmark{12},
R. J. Wainscoat\altaffilmark{12},
C. Waters\altaffilmark{12}
}
\altaffiltext{1} {Kavli Institute for Theoretical Physics, University of California, Santa Barbara, CA 93106, USA}
\altaffiltext{2}{Harvard-Smithsonian Center for Astrophysics, 60 Garden Street, Cambridge, MA 02138, USA} 
\altaffiltext{3}{Department of Astrophysical Sciences, Princeton
University, Princeton, NJ 08544, USA} 
\altaffiltext{4}{Department of Astronomy, University of Illinois at Urbana-Champaign, Urbana, IL 61801, USA } 
\altaffiltext{5}{National Center for Supercomputing Applications, University
of Illinois at Urbana-Champaign, Urbana, IL 61801, USA}

\altaffiltext{6}{Department of Astronomy, University of Washington,
  Box 351580, Seattle, WA 98195, USA}
\altaffiltext{7}{Department of Astronomy and Astrophysics, The Pennsylvania State University, University Park, PA 16802, USA}
\altaffiltext{8}{Institute for Gravitation and the Cosmos, The
  Pennsylvania State University, University Park, PA 16802, USA}
\altaffiltext{9}{Max Planck Institute for Astronomy, K\"{o}nigstuhl 17, D-69117 Heidelberg, Germany}
\altaffiltext{10}{Institute for Applied Computational Science, 
John A. Paulson School of Engineering and Applied Sciences, Harvard
  University, Cambridge, MA 02138 USA}
\altaffiltext{11}{Astrophysics, Imperial College London, Blackett Laboratory, London SW7 2AZ, UK}
\altaffiltext{12}{Institute for Astronomy, University of Hawaii, 2680
  Woodlawn Drive, Honolulu HI 96822, USA}
\altaffiltext{13}{Department of Physics, Durham University, South
  Road, Durham DH1 3LE, UK} 
\altaffiltext{14} {Department of Physics, The Pennsylvania State University, University Park, PA 16802, USA}
\altaffiltext{15}{Einstein Fellow}
\altaffiltext{16}{Alfred P. Sloan Research Fellow}

\title{Detection of Time Lags Between Quasar Continuum Emission Bands 
based on Pan-STARRS Light-curves} 

\begin{abstract}

We study the time lags between the continuum emission of quasars at
different wavelengths, based on more than four years of multi-band
($g$, $r$, $i$, $z$) light-curves in the Pan-STARRS Medium Deep
Fields.  As photons from different bands emerge from 
different radial ranges in the accretion disk, the lags constrain
the sizes of the accretion disks. We select 240 quasars with
redshifts $z \approx 1$ or $z \approx 0.3$ that are relatively emission
line free.  The light curves are sampled from day to month timescales,
which makes it possible to detect lags on the scale of the light crossing time
of the accretion disks. With the code \emph{JAVELIN}, we detect
typical lags of several days in the rest frame between the $g$ band and
the $riz$ bands.  The detected lags are $\sim 2-3$ times larger than the
light crossing time estimated from the standard thin disk model,
consistent with the recently measured lag in NGC5548 and micro-lensing
measurements of quasars. The lags in our sample are found to increase
with increasing luminosity. Furthermore, the increase in lags going
from $g-r$ to $g-i$ and then to $g-z$ is slower than predicted in the thin
disk model, particularly for high luminosity quasars.
The radial temperature profile in the disk must be
different from what is assumed.  We also find evidence that the lags
decrease with increasing line ratios between ultraviolet
\ion{Fe}{2} lines and \ion{Mg}{2}, which may point to changes in the 
accretion disk structure at higher metallicity.

\end{abstract}

\maketitle

\section{Introduction}
The optical/ultraviolet continuum emission from active galactic nuclei
(AGN), particularly at high luminosity, is widely believed to be produced by a
geometrically thin and optically thick accretion disk around the
super-massive black hole (SMBH), where the Eddington ratio for electron
scattering opacity is $\sim 0.01-1$. With the minimal assumption that
the emission is from blackbody radiation with temperature $T$,  for
Eddington luminosity $L_{\rm Edd}$ and emission area $400\pi r_s^2$, 
where $r_s$ is the Schwarzschild radius,
$T\sim \left(L_{\rm Edd}/(400\pi \sigma_r r_s^2)\right)^{1/4}\sim1.2\times 10^5\left(\mbh/10^8\msun\right)^{-1/4}$ K, where
$\mbh$ is the mass of SMBH and $\sigma_r$ is the Stefan-Boltzmann constant. 
This is independent of any accretion disk
model and consistent with the big blue bump in AGN spectra
\citep[][]{Shields1978,KoratkarBlaes1999}.  The standard thin disk
model \citep[][]{ShakuraSunyaev1973} is often used to describe the
accretion disks in quasars. In this model, the effective temperature
$T_{\rm eff}$ changes with radius $R$ as $T_{\rm eff}\propto R^{-3/4}$
for a given black hole mass and accretion rate. Therefore, radiation
at different wavelengths is dominated by emission at different radii.

Because of the large distance of quasars and the small size of their
accretion disks, it is not typically possible to resolve the disk
directly. For a few quasars, microlensing can be used to constrain the
half light radii of the accretion disks
\citep[][]{Morganetal2010,Mosqueraetal2013,Chartasetal2016}.
Variability is another powerful tool to infer the spatial
dimensions of the disks from temporal information
\citep[][]{Lawrence2016}, which can be observed easily.  By studying
the time lags between the light-curves of the continuum emission and
the broad emission lines in AGNs, the size of the broad line region can
be measured, based on the simple assumption that the lag corresponds
to the time photons take to travel from the central black hole to the
broad line region. This well-established
reverberation mapping technique
\citep[][]{BlandfordMckee1982,Peterson1993} has been applied to
many nearby AGNs to study the structure of the broad line region and
estimate black hole masses
\citep[e.g.,][]{Petersonetal2004,Bentzetal2009,Shen2013,Barthetal2015,Shenetal2016}.

In principle, a similar reverberation mapping technique that measures the
lags between the continuum emission in different bands can be used to
constrain the structures of accretion disks. However, the main
challenge is that the expected light crossing time across different
radii in the optical/ultraviolet emission region (particularly the most inner region) 
of the accretion disk 
is much smaller than the lags between the continuum
emission and most broad emission lines.  The light crossing time across 
the expected radii responsible for the continuum emission 
at $\sim 10-100$ Schwarzschild radii in the rest frame of the quasar is $\sim 0.1-1$
day for a $10^8\msun$ SMBH, which means that we need regular
observations with a cadence comparable to or smaller than a day in order to
detect the lags. Accretion disks can also have much longer time
lags between different radii in principle. For example, lags caused by the
propagation of fluctuations in the accretion process happen on the
viscous time scale \citep[][]{Uttleyetal2003,Marshalletal2008}.
However, this is not as clean as the light crossing time scale for the
purpose of constraining accretion disk physics, as detailed modeling
of these long time scale processes is very uncertain.

The short time scale lags between different bands of the continuum
emission have been detected for a few AGNs, including NGC7496
\citep[][]{Wandersetal1997,Collieretal1998}, Markarian 79 \citep[][]{Breedtetal2009}, 
NGC4051 \citep[][]{Breedtetal2010}, NGC3783, MR2251-178 \citep[][]{Liraetal2011}, 
NGC2617 \citep[][]{Shappeeetal2014}, NGC5548 \citep[][]{McHardyetal2014}, 
NGC4395 \citep[][]{McHardyetal2016}, NGC6814 \citep[][]{Troyeretal2016} and upper 
limits of 14
AGNs by \cite{Sergeevetal2005}. Recently, significant lag detections
across a wide range of continuum emission bands have been found for
NGC5548 \citep[][]{Edelsonetal2015,Fausnaughetal2016}.  Most of these
detections show that the short wavelength bands lead the long
wavelength bands, which is usually interpreted as irradiation of the
outer disk by the X-rays produced near the black hole
\citep[][]{Kroliketal1991,Cackettetal2007}. Interestingly, the inferred sizes of the optical 
emitting regions are systematically larger than the 
predicted values from standard thin disk models by factors of $\sim 2-3$
\citep[][]{Lawrence2012,Edelsonetal2015,Fausnaughetal2016}, which is
also consistent with the results based on the microlensing
measurements \citep[][]{Chartasetal2016}.

Theoretically, the standard thin disk model has led to many puzzles
when it is used to describe observations of AGNs in the regime
where it is supposed to apply \citep[e.g.,][]{KoratkarBlaes1999}.  The
radiation pressure dominated inner region of this model, where most of
the continuum radiation is emitted, is thermally unstable
\citep[][]{ShakuraSunyaev1976,Jiangetal2013c}.  However, the expected
large amplitude fluctuations in the thermal time scale caused by the
instability \citep[][]{Janiuketal2002} have never been observed for
most AGNs.  Modifications of the standard thin disk models have been
proposed to explain these discrepancies, including the large temperature
fluctuation model of \cite{DexterAgol2011}, and reprocessing of the
far UV radiation by optically thick clouds \citep{GardnerDone2016}.
Recently, \citet[][]{Jiangetal2016} proposed that the iron opacity 
bump may play an important role
in determining the thermal properties and structure of AGN accretion
disks. The previous continuum lag detections for a few
AGNs are not sufficient to determine the statistical properties of the lags
and test the predictions of these models. The goal
of this paper is to measure the lags between different continuum bands
for a much larger sample of AGNs, which will enable us to quantify the
distributions of the lags and see how the lags change with other
properties of AGNs.

In Section \ref{sec:obsdata}, we describe the data and the sample we
select.  The method we use to measure the lags is described in Section
\ref{sec:lagmethod}.  Our main results are described In Section
\ref{sec:results}. Finally, in Section \ref{sec:discussion}, we
discuss the implications of our results on the understanding of
accretion physics and future work we need.

\section{Observational data and Sample Selection}
\label{sec:obsdata}
\subsection{The Pan-STARRS1 Medium Deep Fields}

We have chosen to study quasars in the Medium Deep Fields of the
Pan-STARRS1 (PS1) survey, because of their depth and large numbers of
imaging epochs.  The PS1 survey used a wide-field $f/4.4$
optical telescope system designed for survey mode operation at the
Haleakala Observatory on the island of Maui in Hawaii. The system,
with 1.8m primary and 0.9m secondary mirrors, produces a 3.3\,deg$^2$
field of view in combination with the PS1 gigapixel camera (GPC1). The
1.4 Gpixel detector is composed of a mosaic of 60 CCD chips each of
4800$\times$4800 pixels with each 10\,$\mu$m pixel spanning 0.258\arcsec\,
on the sky through five main broadband filters denoted as
$g_{\mathrm{P1}}$, $r_{\mathrm{P1}}$, $i_{\mathrm{P1}}$,
$z_{\mathrm{P1}}$, $y_{\mathrm{P1}}$.  The PS1 photometric system is
described in \cite{Tonryetal2012}, and passband shapes are
detailed in \cite{Stubbsetal2010}. The $y_{\mathrm{P1}}$ band data 
typically have fewer points compared with data in 
other bands, and is not used in the following analysis.

While the main 3$\pi$ survey observed three-fourths of the sky north
of $-30^{\circ}$ declination in about a dozen epochs from May 2010 until
March 2014, the Medium Deep Field (MDF) survey of PS1 was designed to
provide deeper exposures with many more epochs in selected fields,
with multiple observations in all 5 filters each season, taken when the airmass 
was lower than 1.3. One MDF cycle starts with 8
$\times$ 113\,s exposures in the $g_{\mathrm{P1}}$ and
$r_{\mathrm{P1}}$ bands on the first night, with 8 $\times$ 240\,s in
the $i_{\mathrm{P1}}$ band the second night, and finally 8 $\times$
240\,s in the $z_{\mathrm{P1}}$ band the third night, before the cycle
recommences.  Any one filter/epoch consists of 8 dithered exposures of
either $8\times 113$s for  $g_{\mathrm{P1}}$ and  $r_{\mathrm{P1}}$ or
$8\times 240$s for the other three, giving nightly stacked images of
904s and 1920s duration.  


The raw science frames exposed with the PS1 telescope were reduced by
the PS1 Image Processing Pipeline (IPP) conducting standard procedures
of image calibration, source detection, astrometry, and photometry.
We use an updated version of the ``ubercalibrated" PS1 data from
\cite{Schlaflyetal2012}, which includes the PS1 data up
through PV1 (using PV1 of the PS1 pipeline) and is calibrated
absolutely to 0.02 magnitudes or better. This database excludes
detections flagged by PS1 as cosmic rays, edge effects and other
defects.  We consider only 9 PS1 MD fields (1, and 3 -- 10) that
overlap the SDSS footprint.  The median number of PS1 epochs
is 284, 340, 406, 445, 179 in the $g$, $r$, $i$, $z$, $y$ filters, respectively,
and median magnitudes are in the range $16<i_{\mathrm{P1}}<21.5$. 
As determined by the analysis of non-variable stars in Morganson et
al. (2015),  the magnitude uncertainties delivered by the PS1 pipeline
have been inflated, by 1.387, 1.327, 1.249, 1.228, and
1.170 for {\em g, r, i, z, y}, respectively.



\subsection{The Parent Quasar Sample}

We began by searching for all objects within 1.5$^{\circ}$ of each MDF
central coordinate that have been observed and spectroscopically
classified as a quasar within SDSS Data Release 10 (part of SDSS-III;
\citealt{Eisensteinetal2011,Parisetal2014}) by querying the CasJobs
data server.  This yielded 2421 unique quasars.

We also included quasars identified in two ancillary pilot programs
using the multi-fiber Baryon Oscillation Spectroscopic Survey (BOSS)
spectrograph within SDSS-III \citep[][]{Dawsonetal2013}.  
The SDSS-III program approved
spectroscopy of variables and X-ray source counterparts within MD01
and MD03, as a pilot study for two spectroscopic subprograms of eBOSS
proposed (and currently underway) for SDSS-IV: SPectroscopic
IDentification of E-Rosita Sources (SPIDERS) is for X-ray source
follow-up, and the Time Domain Spectroscopic Survey (TDSS) is for
classification and study of photometric variables.  Both programs are
briefly described in \cite{Alametal2015}.  Despite the reference to
eROSITA, SPIDERS sources actually were selected optical counterparts to
catalogued ROSAT and/or XMM-{\em Newton} X-ray sources. The TDSS pilot
variable candidates were selected from early PS1 MDF ubercalibrated
(Schlafly et al. 2012) photometry, when typically about a third of the
final number of epochs was available.  TDSS pilot candidate variable
priorities were assigned from a weighted sum of variability features
$R_{CS}$, SNR and the median SDSS-PS1 magnitude difference across the
{\em griz} bands.  SNR in each filter is defined as the ratio of the
75\%-25\% magnitude quartiles divided by the median magnitude error
derived from the full light-curve. $R_{CS}$ is the range of a
cumulative sum from \cite{Ellaway1978}.  These statistical features and
their use are described in \cite{Kimetal2011}.

%
%
%

For MD01, plate number 6369 was observed for a total of 75min on Oct
13, 2012 (MJD 56217), while for MD03, plate 6369 was observed for a
total of 60min on Dec 23, 2012 (MJD 56284).  We include all the
quasars spectroscopically identified in the MD01 and MD03 pilot
campaigns, both by the SDSS pipeline and as confirmed from our own visual
inspection.  We note that, given the X-ray and optical photometric
variability selection, the quasar samples in these 2 fields therefore
have a somewhat different set of selection biases than the usual SDSS
optical color quasar selection algorithms.  We further include all 991
spectroscopic quasars known within MD07, of which 849 are targeted in
an ongoing campaign of repeated spectroscopy for reverberation mapping
there \citep[][]{Shenetal2015}.   Before completion of this paper,
all the included quasar spectra have become available to the public
via SDSS Data Release 12.


Since we are interested in understanding the physics of accretion in
our sample, we seek to analyze only quasars for which spectroscopy
allows a reasonable estimate of the SMBH mass, using the single-epoch
virial method described by \cite{Shenetal2011}.

Across all 9 MDFs, we identified 3178 unique quasars that have ${M_{\rm BH}}$
fits and PS1 MDF light-curves that pass all our criteria.  
The ${M_{\rm BH}}$ measurements have been performed with different
broad emission line fits, depending on which lines are available in the SDSS
spectrum as a
function of quasar redshift.  For $z<0.76$, we choose to use H$\beta$ and
$\lambda_0\,5100$\AA\ continuum, with $k_{\rm Bol}=9.26$.
For $0.76<z<2.1$, we use Mg\,II and $\lambda_0\,3900$\AA, continuum
with $k_{\rm Bol}=5.15$.  For $2.1<z<3.18$, we use C\,IV and 
$\lambda_0\,1350$\AA, continuum, with $k_{\rm Bol}=3.81$.
The fit errors are generally lower for spectra with higher
signal-to-noise ratio (S/N), which is markedly higher (typically
$\sim$30/pixel) within MD07, because we fit spectra of 32 co-added
observation epochs obtained for the reverberation mapping campaign \citep[][]{Shenetal2015}.

%
%
%
%

\subsection{The Final Quasar Sample}

Broad emission line fluxes are known to vary in response to continuum
flux variations, but with a time delay related to their physical
distance from the continuum-emitting region.  While this effect is
exactly what allows the reverberation mapping method, we are seeking
to detect shorter delays that may occur {\em between}
continuum-emitting regions. The existence of broad emission lines 
in the broad bands may contaminate the possible lags between 
continuum emissions in standard lag analysis \citep[][]{CheloucheZucker2013}.
Therefore, from the larger sample of
quasars with SDSS spectra, black hole mass
estimates, and PS1 light-curves, we select for further analysis those
quasars in redshift ranges where broad emission lines present the least
contribution within the PS1 broad-band filter transmission curves 
(particularly $g$ band),
yielding 51 quasars with $0.16<z<0.42$ and 189 with $0.95<z<1.1$.  
All the quasars in our sample and their properties are summarized in 
Table \ref{table:sample}.

\begin{figure}[h]
\centering
\includegraphics[width=1.0\hsize]{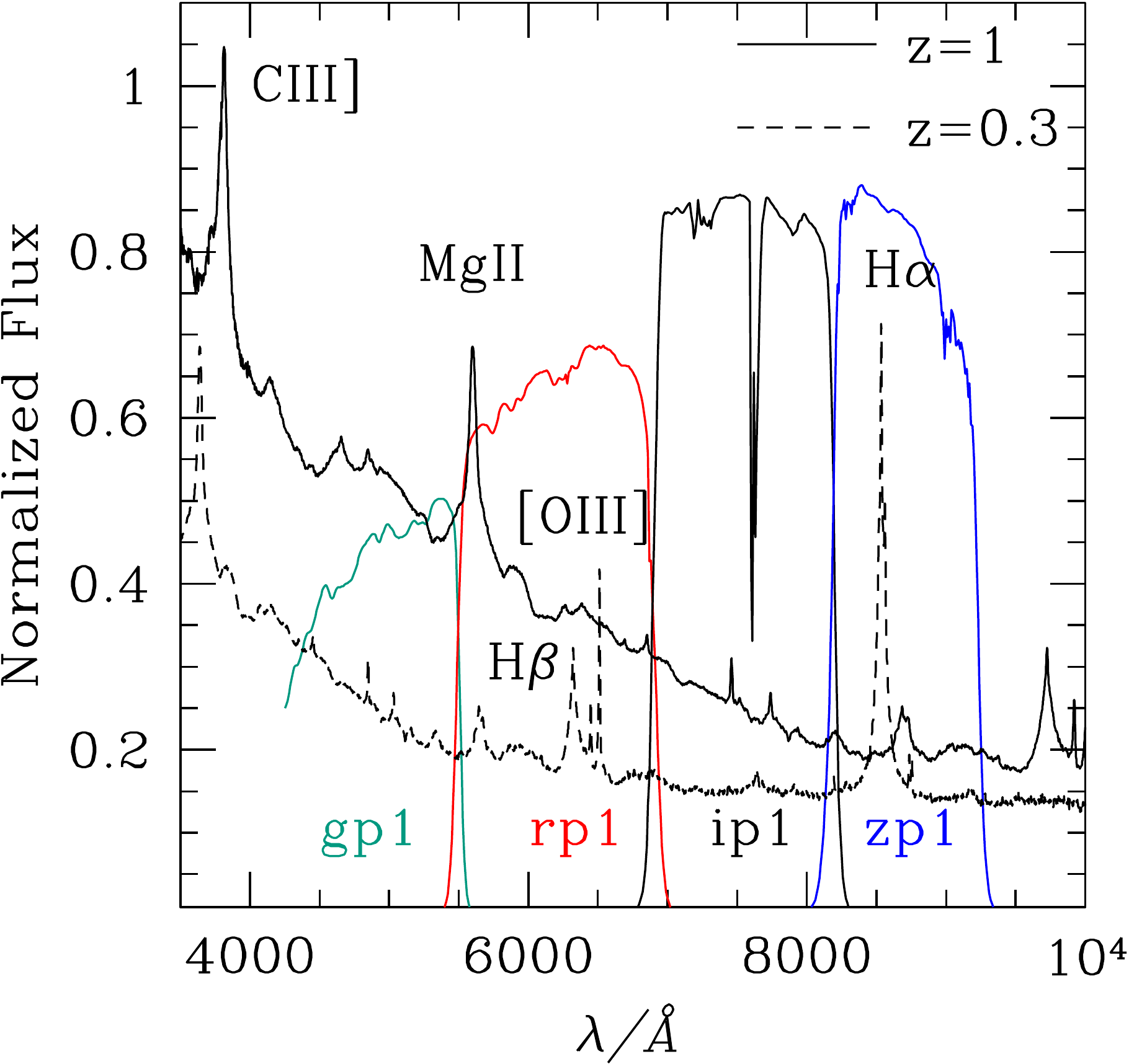}
\caption{Pan-STARRS broad band filters 
as given by \cite{Tonryetal2012} and typical quasar 
spectrum at redshift $1$ and $0.3$ from \cite{Telferetal2002}. }
\label{filterspectrum}
\end{figure}


\begin{figure*}[h]
\centering
\includegraphics[width=0.82\hsize]{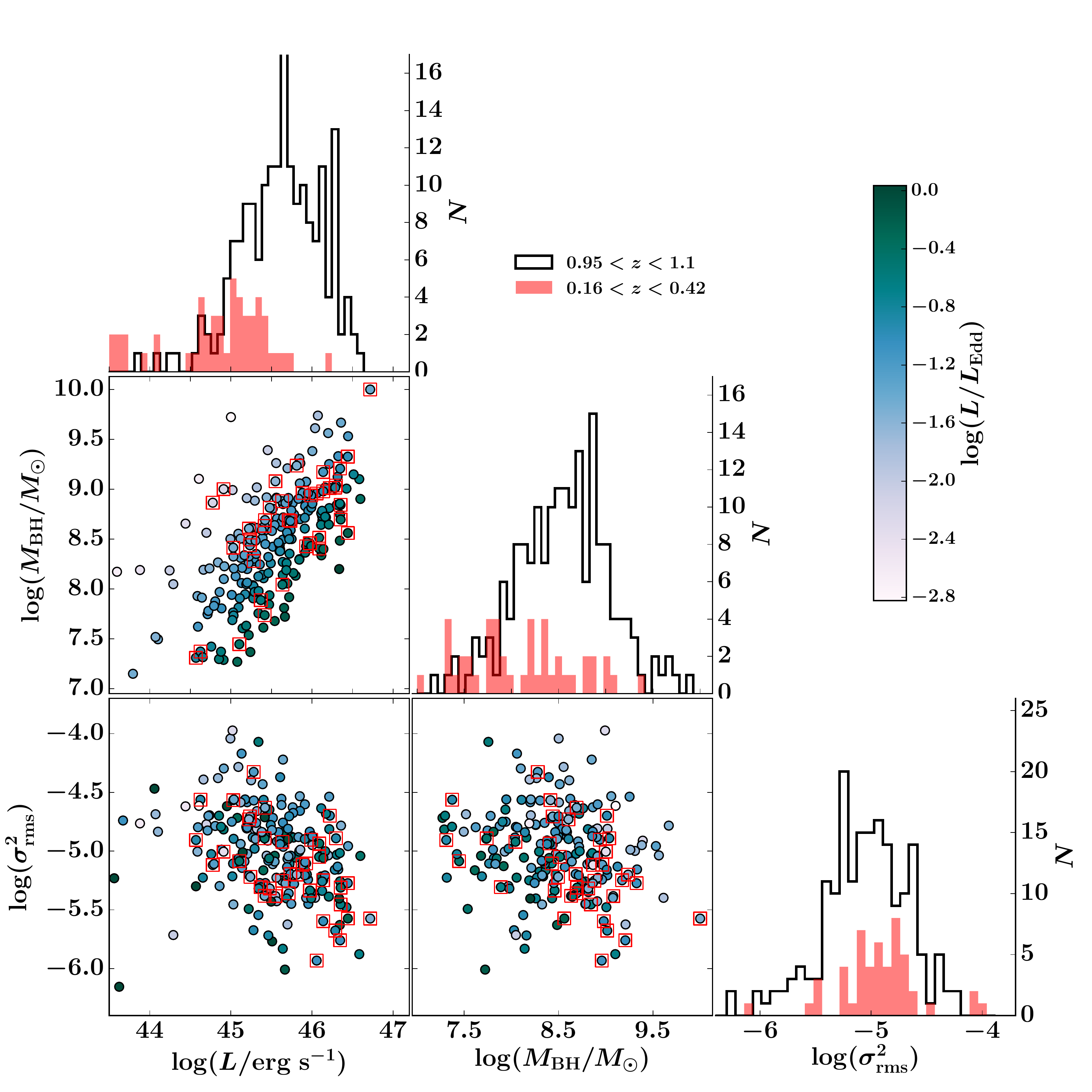}
\caption{Distributions of the estimated bolometric luminosity $L$, black hole
mass $\mbh$ and normalized $g$ band excess variance 
$\sigma^2_{\rm rms}$ for the $200$ quasars with detected lags. Each quasar is
color coded by Eddington ratio. The $39$ quasars with significant
detections (subsample \emph{cLD}) are labeled by the open red
squares. The diagonal panels are the histograms of $L$, $\mbh$ and
$\sigma^2_{\rm rms}$ in the two redshift bins.}
\label{SampleProperty}
\end{figure*}

Distributions of the estimated luminosity, black hole mass and
Eddington ratio for our sample are shown in Figure
\ref{SampleProperty}. The sample covers a luminosity range from 
$\sim 3\times 10^{43}$ to $10^{47}$ erg s$^{-1}$ and the estimated black hole masses
range from
$10^7$ to $10^{10}\ M_{\odot}$.  As expected, the low redshift quasars
also typically have lower luminosity.

One common way to quantify the level of variability in
the light-curves is the normalized excess variance  
$\sigma_{\rm rms}^2$ \citep[][]{Nandraetal1997,Vaughanetal2003}, which is basically the standard
deviation of the measurement error corrected flux scaled by the mean flux and it is calculated
according to Equation 1 of \cite{Simmetal2016}. 
In order to minimize the effects of light curve gaps, we calculate the excess variance 
in each season and take the average value of $\sigma_{\rm rms}^2$ over the four seasons. 

The distributions of
$\sigma_{\rm rms}^2$ for the $g$ band light-curve of our sample are
also shown in Figure \ref{SampleProperty}.   The normalized excess
variance shows an anti-correlation with luminosity and with black
hole mass, which can also be captured by a single anti-correlation
between $\sigma_{\rm rms}^2$ and Eddington ratio $L/L_{\rm Edd}$, with
$\sigma_{\rm rms}^2\propto \left(L/L_{\rm Edd}\right)^{-0.20\pm0.11}$. 
The Pearson and Spearman $p$-values of  the anti-correlation, which are the probability that 
$\sigma_{\rm rms}^2$ is not correlated with $L/L_{\rm Edd}$, 
are $3.0\times 10^{-4}$ and $3.8\times 10^{-4}$ respectively.  A
similar anti-correlation between the overall (long-term) variability
amplitude and $L/L_{\rm Edd}$ has been found by other studies
\citep[][]{Wilhiteetal2008,Baueretal2009,Aietal2010,MacLeodetal2010,Pontietal2012}.

\begin{table*}
\centering
\caption{Sample Summary}
 \label{table:sample}
\scalebox{0.9}{
\begin{tabular}{ccccccccccc}
\hline \hline
                      &   RA     & DEC  &        & $m_g$& $m_r$ & $m_i$  & $m_z$   & 	 $\log L$   &  $\log\mbh$	&\\
SDSS Name  &  (Deg) & (Deg) & $z$ & (mag) &  (mag)  &  (mag) &   (mag)   &	(erg/s)	&  $\msun$      &  $\sigma^2_{\rm rms}$       \\
 \hline
 J022303.23-025713.8 & 35.7635 & -2.9539  & 0.41 & 19.36$\pm$0.01 & 19.39$\pm$0.01 & 19.26$\pm$0.02 & 18.89$\pm$0.04 & 45.09 & 7.81 & 8.68e-06  \\ 
J022115.53-025843.4 & 35.3147 & -2.9787  & 0.99 & 20.91$\pm$0.03 & 20.85$\pm$0.05 & 20.78$\pm$0.06 & 20.70$\pm$0.19 & 45.20 & 8.20 & 2.61e-05  \\ 
J021809.24-035848.7 & 34.5385 & -3.9802  & 0.98 & 21.95$\pm$0.08 & 21.78$\pm$0.09 & 21.49$\pm$0.10 & 21.47$\pm$0.40 & 45.08 & 7.27 & 1.92e-05  \\ 
J021800.49-040649.2 & 34.5021 & -4.1137  & 1.04 & 21.15$\pm$0.04 & 20.86$\pm$0.04 & 21.00$\pm$0.07 & 20.43$\pm$0.18 & 45.17 & 7.95 & 2.75e-05  \\ 
J022616.01-030537.0 & 36.5667 & -3.0936  & 0.99 & 21.41$\pm$0.05 & 21.10$\pm$0.06 & 21.32$\pm$0.10 & 20.77$\pm$0.23 & 44.95 & 0.00 & 2.41e-05  \\ 
J022808.90-035845.3 & 37.0371 & -3.9792  & 0.99 & 21.93$\pm$0.07 & 21.55$\pm$0.08 & 21.40$\pm$0.10 & 21.92$\pm$0.53 & 44.10 & 7.50 & 1.46e-05  \\ 
J022521.25-032628.5 & 36.3386 & -3.4413  & 1.07 & 20.68$\pm$0.03 & 20.17$\pm$0.03 & 20.42$\pm$0.05 & 20.09$\pm$0.10 & 45.39 & 8.51 & 3.71e-05  \\ 
J022659.82-035015.0 & 36.7493 & -3.8375  & 0.29 & 20.97$\pm$0.03 & 20.20$\pm$0.03 & 19.99$\pm$0.03 & 19.49$\pm$0.07 & 44.57 & 7.31 & 1.24e-05  \\ 
\hline
\end{tabular}}
\begin{tablenotes}
\item Note: This table is available in its entirety in a machine-readable form in the online journal. 
A portion is shown here for guidance regarding its form and content.
\end{tablenotes}
\end{table*}

\begin{table}
\centering
\caption{Sample Statistics}
 \label{table:numbers}
\begin{tabular}{cccc}
\hline \hline
Name  &  No. & $0.95<z<1.1$ & $0.16<z<0.42$ \\
 \hline
Initial Sample & 240 & 189 & 51 \\
\emph{noLag} & 40 & 29 & 11 \\
\emph{iLD} & 200 & 160 & 40 \\
\emph{cLD} & 39 & 34  & 5 \\
\hline
\end{tabular}
\begin{tablenotes}
\item Note: {\sf JAVELIN} does not detect any 
lag signal from the subsample named \emph{noLag}. 
The subsample \emph{iLD} has lag detections while 
\emph{cLD} is the subsample with single significantly peaked lags 
and lags increase with increasing wavelength differences. 
\end{tablenotes}
\end{table}

\section{Lag Analysis Method}
\label{sec:lagmethod}

\begin{figure*}[htp]
\centering
\includegraphics[width=0.48\hsize]{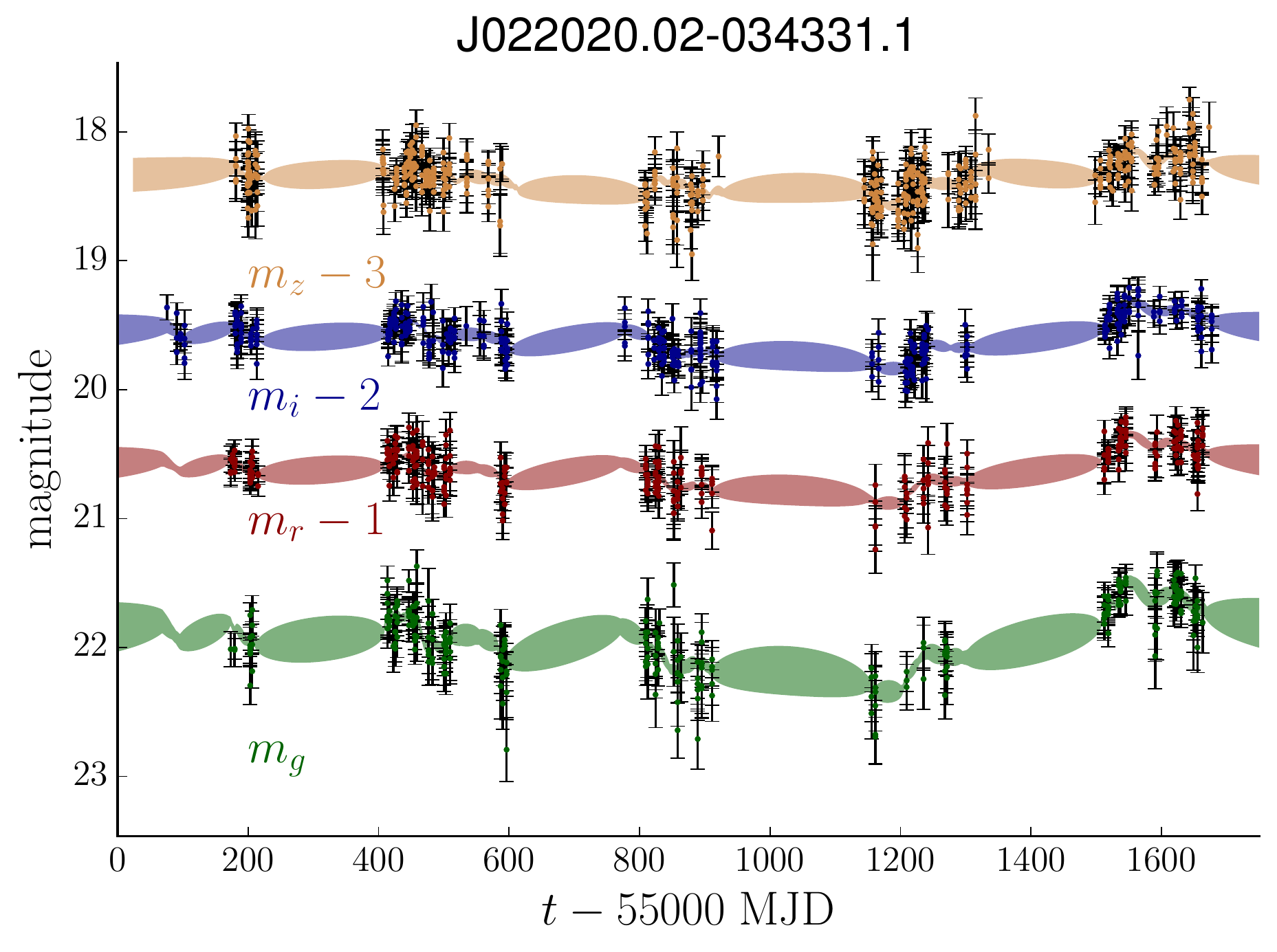}
\includegraphics[width=0.48\hsize]{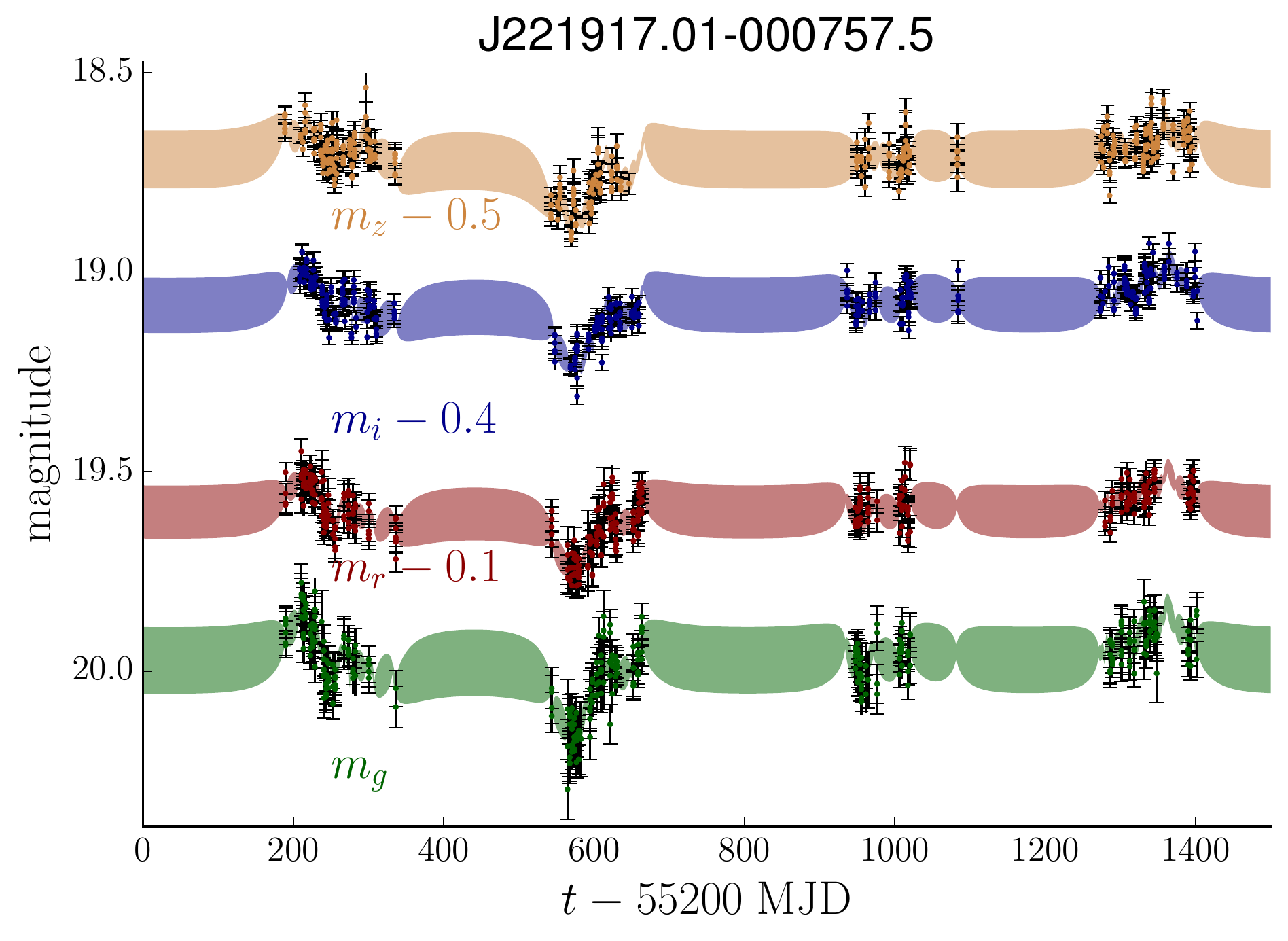}
\caption{Example light-curves in the $g$, $r$, $i$ 
and $z$ bands for the two sample quasars J022020.02-034331.1 in MD01
and J221917.01-000757.5 in MD09. The shaded region 
is the weighted mean of {\sf JAVELIN} light-curves that are consistent 
with the data and the $1\sigma$ dispersion of those light-curves. 
Light-curves for the full sample with {\sf JAVELIN} fits 
are available online.}
\label{LCExample}
\end{figure*}

Cross-correlation is the traditional method to calculate lags between
two light-curves \citep[e.g.,][]{Petersonetal1998}. This method works
well for well-sampled light-curves and it does not assume any
model for the continuum light-curve. However, cross-correlation
interpolates between the data points, so it may not be easy to pick out
lag signals from light-curves with large and irregular gaps. This
is the case for our data.  Example light-curves for two quasars in
MD01 and MD03 are shown in Figure \ref{LCExample}. The Pan-STARRS
light-curves typically have cadences varying from a day to a few
months with large gaps between seasons. For this reason, we 
use {\sf JAVELIN} \citep[][]{Zuetal2011,Zuetal2013} to calculate the
lags between different bands.

\subsection{Fitting Procedure with {\sf JAVELIN}}

Because quasar variability is found to be acceptably described by the
damped random-walk (DRW) model
\citep[][]{Kellyetal2009,Kozlowskietal2010,MacLeodetal2010},  
which is a first-order continuous autoregressive model,
{\sf JAVELIN} first fits a DRW model to the $g$ band data to get the
variability amplitude $\sigma$ and the damping time scale $\tau$. The
model is then smoothed and shifted to fit the $r$ and $i$ bands simultaneously. The
lags between the $g$-$r$ and $g$-$i$ bands are determined when best
fits for the three bands are achieved.  We have also confirmed
that if we fit the $g$-$r$ and $g$-$i$ bands independently, the
signals we find are consistent with the previous case within the
uncertainty of the peaks.
We repeat the process to calculate the lag between the $g$ and $z$
bands. For each quasar, we carry out $90000$ Monte Carlo Markov Chain
(MCMC) burn-in and 
sampling iterations to estimate the probability distributions of lags and
other parameters of DRW models. In our sample, there are $40$ quasars
(labeled as \emph{noLag}) where {\sf JAVELIN} cannot find a
significant lags in this approach (see discussion section
\ref{sec:discussion}).  This leaves 37 quasars between $z=0.16$ and
$0.42$ and 163 between $z=0.95$ and $1.1$ with detected signals
(labeled as \emph{iLD}), which will be the focus of our analysis.

Notice that although {\sf JAVELIN} was originally developed to
calculate lags between the continuum and lines, we can easily
replace the lines with continuum emission in different bands, since
the different continuum bands should, at least to first order approximation, 
follow the same variability
process but with a delay.  In fact, {\sf JAVELIN} has been
successfully used to calculate continuum-continuum lags in NGC5548
\citep[][]{Fausnaughetal2016}, where it agrees with the cross-correlation
methods well for these well-sampled light-curves.

\subsection{Testing {\sf JAVELIN}}
\label{sec:test}

The irregular cadence and the fact that not all bands are observed
simultaneously can introduce artificial lag signals in principle. We
have done a series of experiments to test the effects of the cadence
on the lag signals detected by {\sf JAVELIN}.

\begin{figure}[h]
\centering
\includegraphics[width=1.0\hsize]{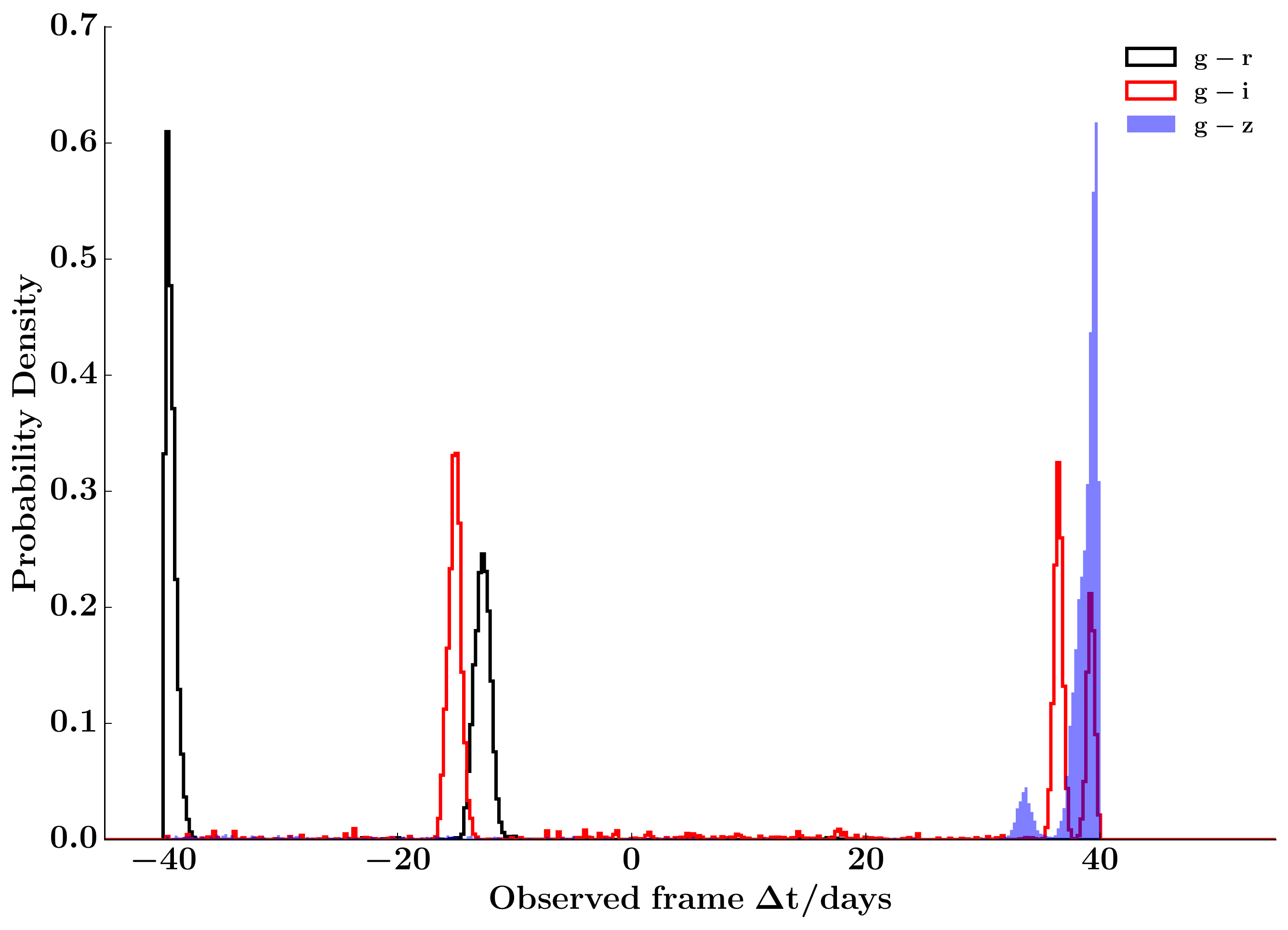}
\includegraphics[width=1.0\hsize]{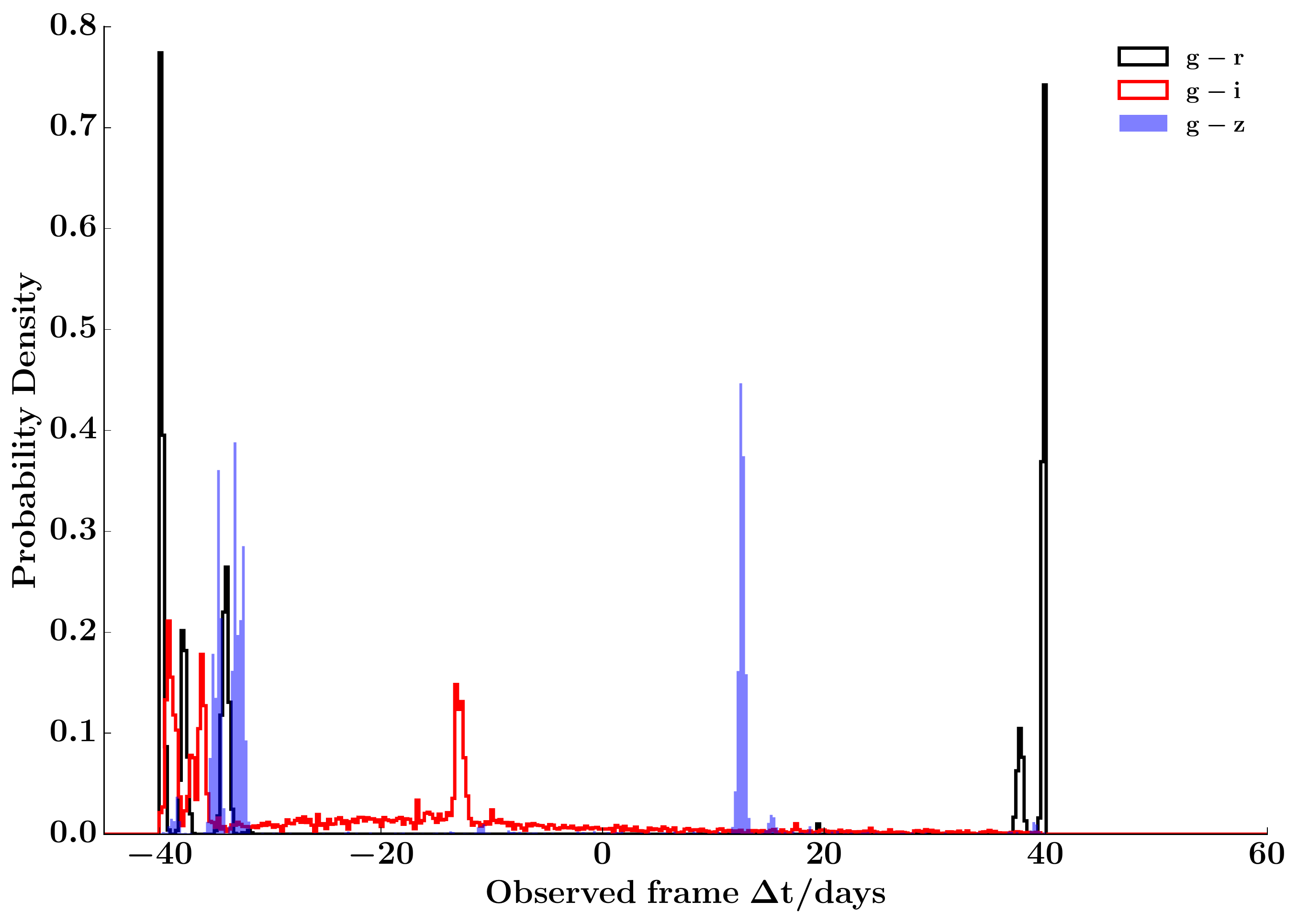}
\caption{Probability density distributions of lags in the observed frame 
between the 
actual $g$ band data and randomly generated mock light 
curves in $r$, $i$ and $z$ bands as described in Section \ref{sec:test}. 
The top panel is for quasar  
J022020.02-034331.1 in MD01 while the bottom panel is for 
quasar J221917.01-000757.5 in MD09.}
\label{testrandom}
\end{figure}

The first set of experiments gives {\sf JAVELIN} pairs of
light-curves without any lag signals. We calculate the mean magnitude
$\mu_0$ and standard deviation $\sigma_m$ in the $r$, $i$, and $z$ bands.
We take the time of each data point in these bands but assign a
magnitude $\mu_0+\sigma_m s$, where $s$ is a normally distributed
random variable. The error on each data point is constructed in
the same way based on the mean and standard deviation of the error
bars of the original data. In this way, we construct mock light-curves
in the $r$, $i$ and $z$ bands using the actual cadence but
uncorrelated magnitudes. We feed {\sf JAVELIN} the actual $g$ band
data and these mock light-curves to calculate the lags following the
same procedures as we have described before. The resulting probability
density distribution is usually uniformly distributed over all the
possible lags.  However, some common spurious lags show up. Two
examples are shown in Figure \ref{testrandom}.  These lags are usually
located at special locations for different quasars in the same field
such as $-40, -15, 15, 40$ days, which are likely caused by the
cadence of the observations as similar values also show up in 
the probability density distribution of cadences. 
This may also explain similar lag
signals we see in some quasars when we calculate the lags using the
actual data. We emphasize that we never see any artificial signals
around time scales of a few days in this experiment. To test the effects of 
different noise models, we have also constructed four independent 
DRW light curves, which are mapped to the actual MJDs in $g,r,i,z-$bands.  
Error bars of the original data are assigned to these mock light-curves. 
We find very similar results as in the case when we use white noise mock 
light-curves. 

\begin{figure}[h]
\centering
\includegraphics[width=1.0\hsize]{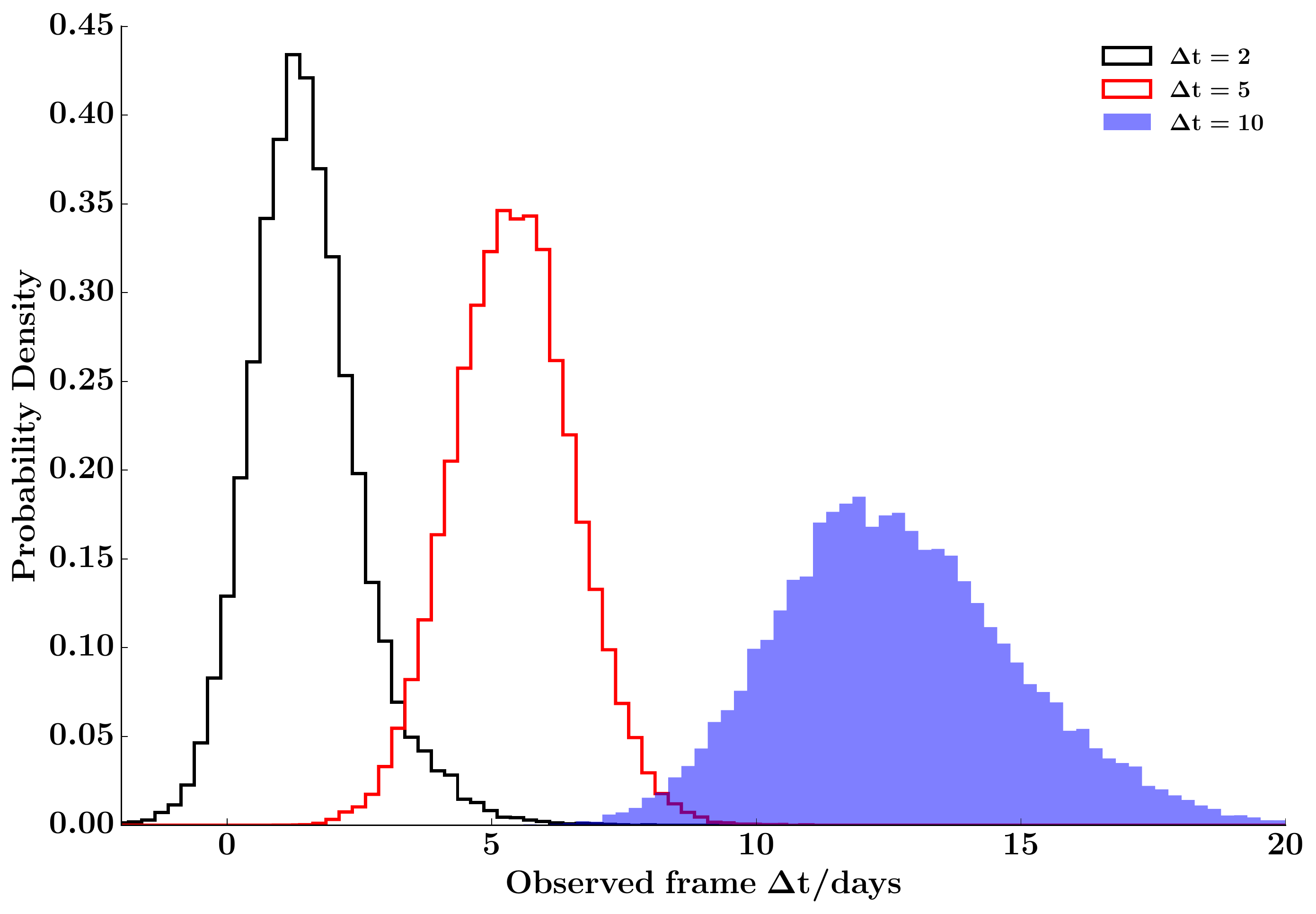}
\caption{Probability density distributions of the lags in the observed frame 
as calculated by {\sf JAVELIN} between the actual $g$ band data and 
the shifted light-curves with $2$, $5$ and $10$ days as described in  
Section \ref{sec:test}. This example is for quasar J022020.02-034331.1 
in MD01. }
\label{testlag}
\end{figure}

To see whether {\sf JAVELIN} is able to detect genuine lags based on
the Pan-STARRS light-curves, we generate mock light-curves with
specified lags. We first take the actual $g-$band light-curve and add
$2,5$ and $10$ days respectively to the time of each data point to
make three mock light-curves. The magnitude and error on each data
point in the mock light-curves are the same as in the original data. These 
mock light-curves are then mapped to the actual MJDs in $r$ band via 
linear interpolation. 
This effectively introduces $2$, $5$ and $10$ days' lags (in the observed
frame) in the mock light-curves. We provide {\sf JAVELIN} with the
actual $g-$band data and the mock light-curves as three data sets to
calculate the lags.  Probability distributions of the lags in this
experiment for quasar J022020.02-034331.1 are shown in Figure
\ref{testlag}.  The peaks of the calculated lags from {\sf JAVELIN}
are located at the locations of the input signals, except for the case with 
a lag of $10$ days, where the probability density distribution peaks at 
$12$ days with a standard deviation $2.2$ days. The large uncertainty and 
offset in this case are likely due to the interplay between the input signal and 
the light curve cadences. 
The lag distributions are also typically broader for quasars with larger magnitude
uncertainties.  
This experiment shows that {\sf JAVELIN} is able to
pick out lags as short as 2 days (in the observed frame) even given
the irregular cadence of the Pan-STARRS light-curves.
 
 \begin{figure*}[h]
\centering
\includegraphics[width=0.33\hsize]{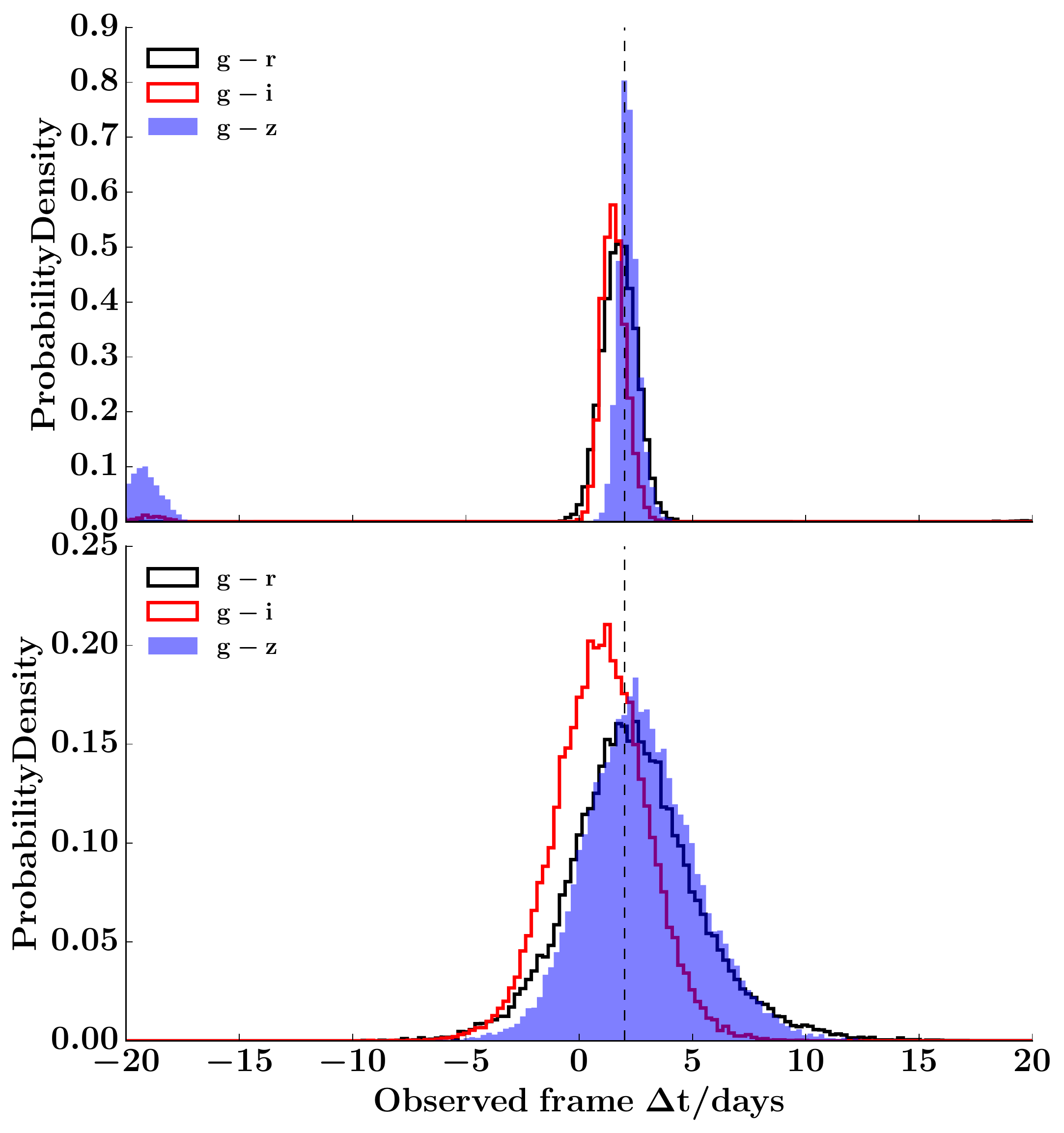}
\includegraphics[width=0.33\hsize]{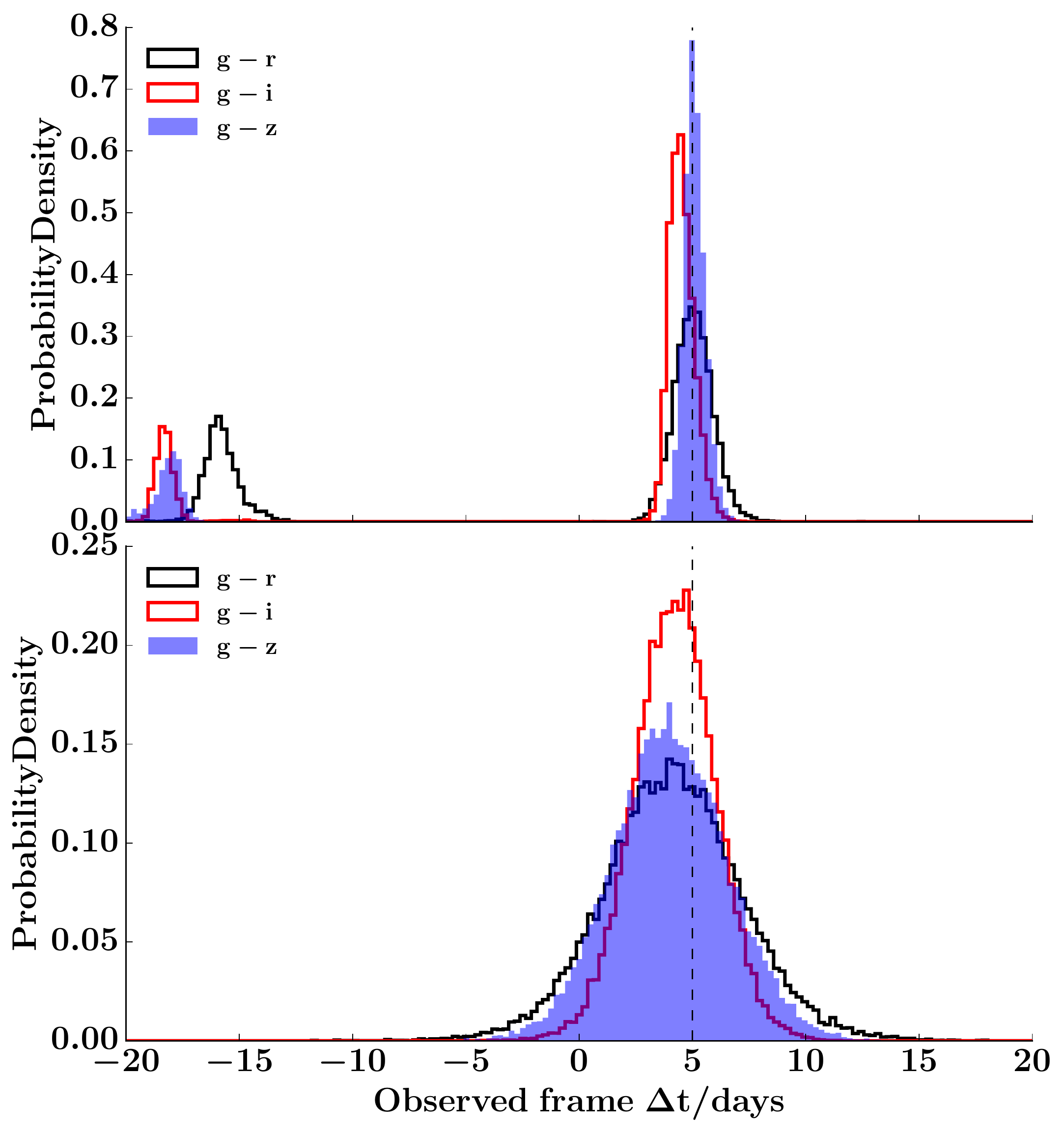}
\includegraphics[width=0.33\hsize]{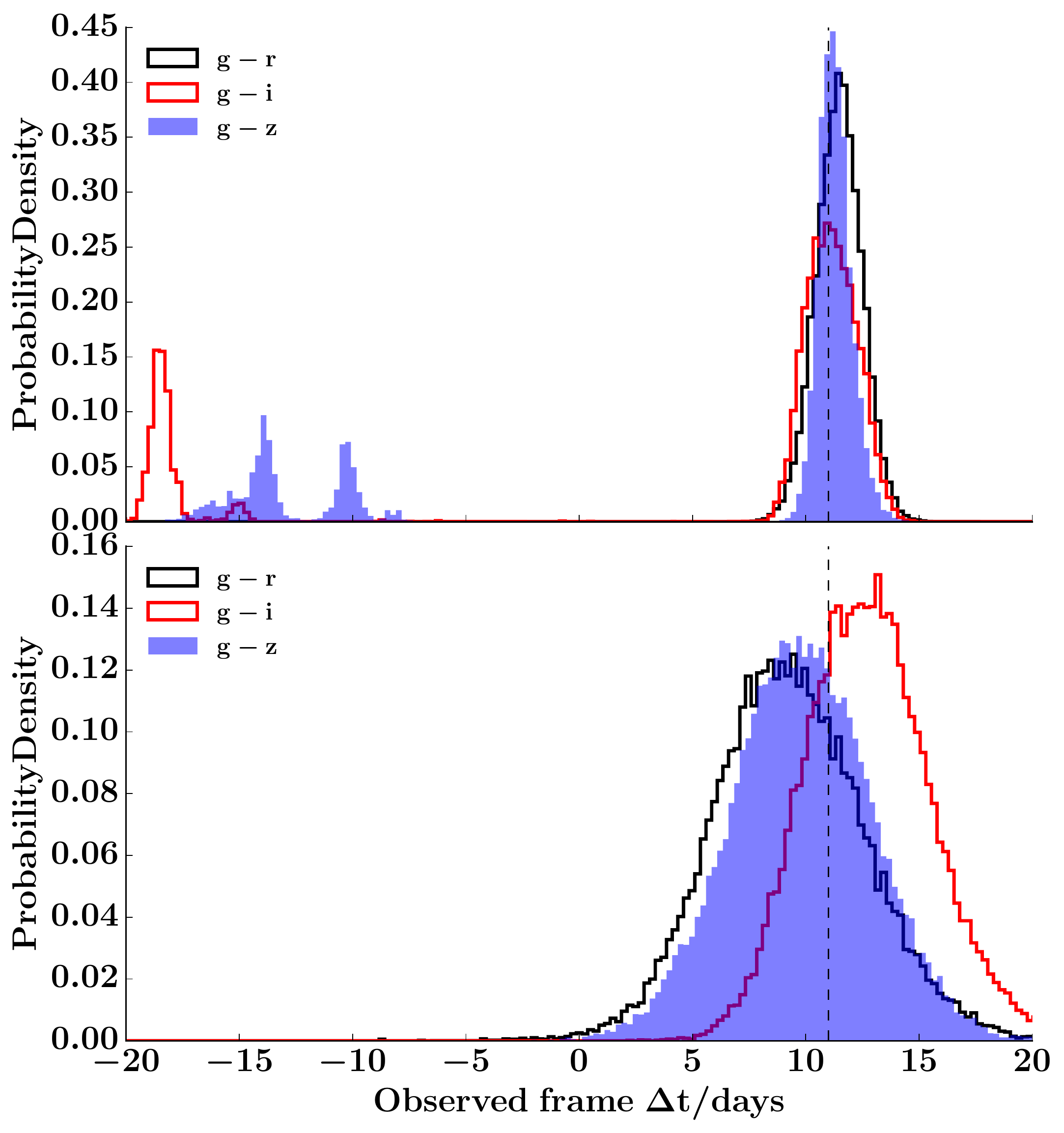}
\caption{
Probability density distributions of lags in the observed frame
between the mock $g$ band light-curve and shifted and resampled mock
$r$, $i$ and $z$ band light-curves. The mock light-curves are
constructed based on a high cadence damped random-walk light-curve as
described in Section \ref{sec:test}. From left to right, they are
cases when the mock $r$, $i$ and $z$ band light-curves are shifted by
$2$, $5$ and $11$ days respectively as indicated by the vertical
dashed lines.  The bottom panels are for quasar J022020.02-034331.1 in
MD01 while the top panels are for quasar J221917.01-000757.5 in
MD09. The mean $g-$band magnitude uncertainties for
J221917.01-000757.5 and J022020.02-034331.1 are $0.01$ and $0.09$ mag
respectively, which explains why the probability density distributions
in the bottom panel are much broader than the distributions in the top
panels. }
\label{testbandlag}
\end{figure*}

We also perform similar experiments with mock light curves for
different bands based on the DRW model. We generate a DRW light curve using the best-fit
parameters (the damping time scale $\tau$ and variation amplitude
$\sigma$) as returned by {\sf JAVELIN} for the $g-$band data. This
light curve is uniformly sampled with cadence $0.05$ day and covers
the full time interval of the $g$, $r$, $i$ and $z-$band data. We
generate mock light curves by interpolating the high-cadence
light curve at the observation times of the data points in each
band. In this way, we get the same light curve sampled at different
MJDs in the four bands.  Error bars on the mock light curve magnitudes
in each band are taken to be the mean error bar of the actual
light curve in the same band. We have also tried using the actual
errors from the original light curves, which do not show any
difference.  We then shift the high cadence light-curve by $1$ to $13$
days and map to $r$, $i$ and $z$ bands.  We use {\sf JAVELIN} to
calculate the lags between the $g$ band and $13$ shifted light-curves
resampled in the $r$, $i$ and $z$ bands. Figure \ref{testbandlag}
shows examples of the calculated lags for two quasars. {\sf JAVELIN}
recovers the lags we inserted between the $g-$band and the other
bands in all cases, although the uncertainty is clearly larger
compared with the previous experiment. In some cases, as shown in the
top panel of Figure \ref{testbandlag}, when the input lag is around
$10$ days, artificial negative lags around $-15$ to $-20$ days show
up. This is likely caused by the combination of the input signal and the
cadence, as these artificial signals only show up with input signals
around $10$ days and they are usually located at $-10$ to $-20$
days. We have also checked that we never see any spurious lags 
around time scales of a few days in this experiment. 

The mock light curves generated in the above experiments correspond to the case with a 
$\delta$ transfer function. In order to test the effects of a finite width in the transfer function, 
we generate new mock light curves by convolutions between the high cadence DRW light 
curve and a log-normal transfer function \citep[][]{Starkeyetal2016} 
$f(\Delta t)=\exp\left[-(\log(\Delta t)-\mu)^2/(2\sigma^2)\right]/\left(\left(2\pi\right)^{1/2}\sigma \Delta t\right)$. 
This effectively introduces a mean lag $\Delta t=\exp[\mu+\sigma^2/2]$ in the mock light curves while 
$\sigma$ determines the width of the transfer function. We have tried $\sigma=0.1, 0.2, 0.5$ and in each 
case we generate mock light curves with $\mu=\log(2)+\sigma^2$, $\log(5)+\sigma^2, \log(10)+\sigma^2$. 
These mock light curves are mapped to $r$, $i$ and $z$ bands and we feed them to {\sf JAVELIN} to 
calculate the lags with respect to the $g$ band mock light curve. We carry out this experiment for the 
same two quasars as in Figure \ref{testbandlag}.  
{\sf JAVELIN} is still able to recover the mean lag values as in Figure \ref{testbandlag}. However, 
the uncertainty is significantly increased with larger $\sigma$. In the case of $\sigma=0.5$, FWHM of the 
probability density distribution is increased by a factor of $\sim 2-2.5$ compared with the 
results from the previous experiment.

The damping time scale $\tau$ we get by fitting DRW 
models to the light curves of quasars in our sample varies 
from $30$ to $500$ days. We have also tried the experiment of  
forcing $\tau$ to be larger than $200$ days in {\sf JAVELIN}, which is 
the typical value found for most quasars \citep[][]{MacLeodetal2010}, 
and we find almost identical lag signals. This demonstrates that the DRW 
parameters we get from {\sf JAVELIN} may not be robust, but 
the lags we aim to detect are.

\begin{figure*}[h]
\centering
\includegraphics[width=0.33\hsize]{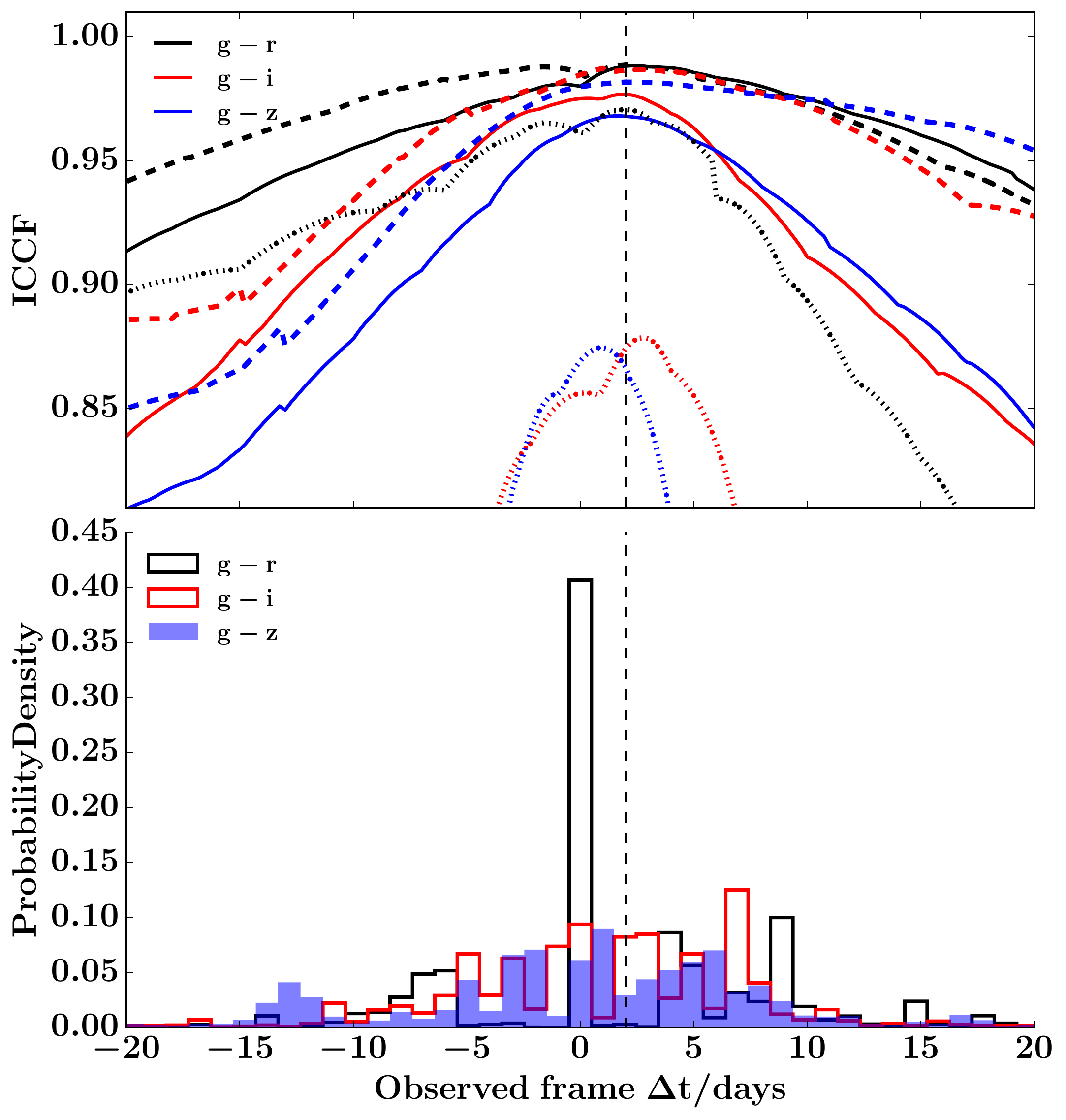}
\includegraphics[width=0.33\hsize]{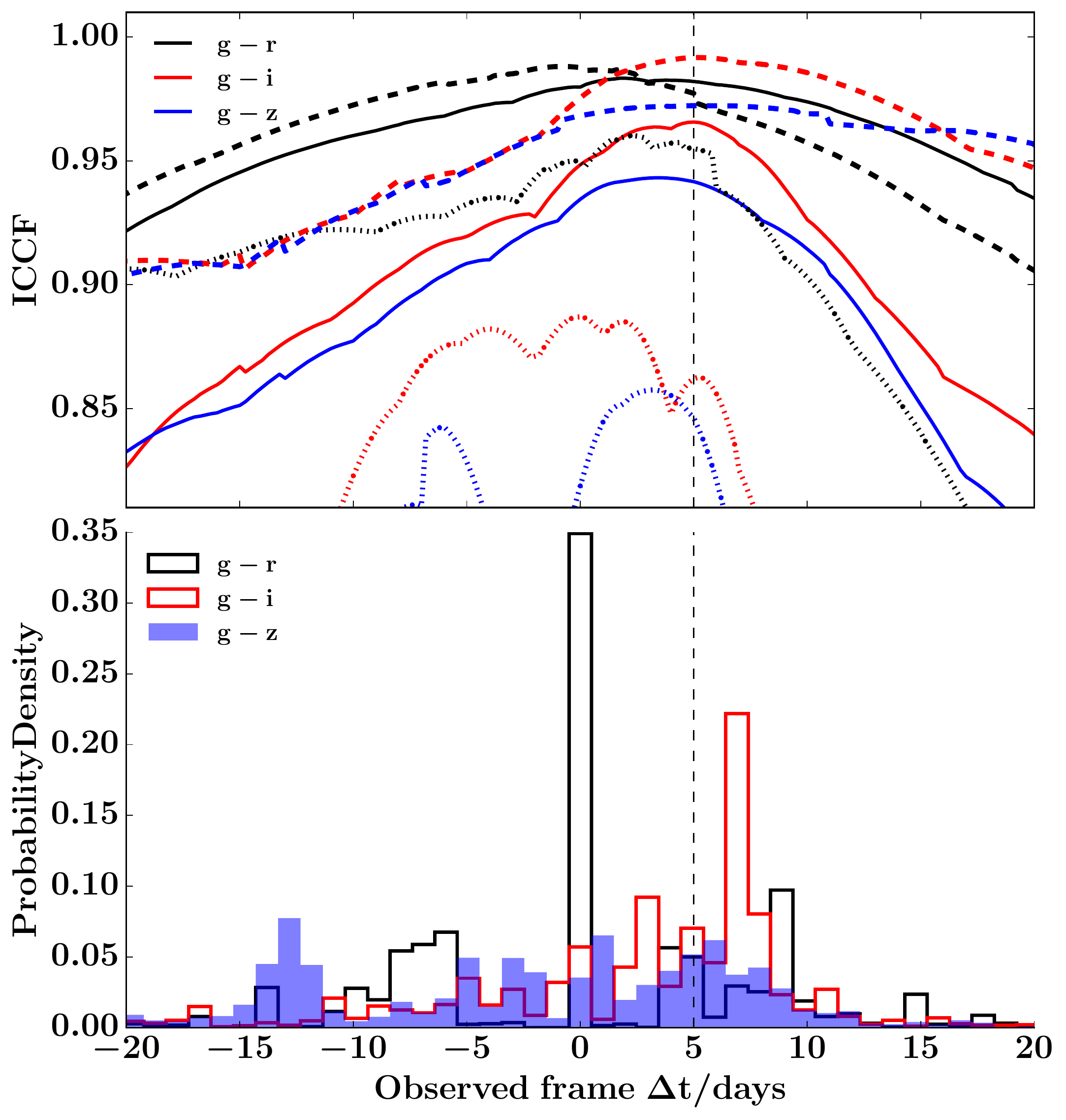}
\includegraphics[width=0.33\hsize]{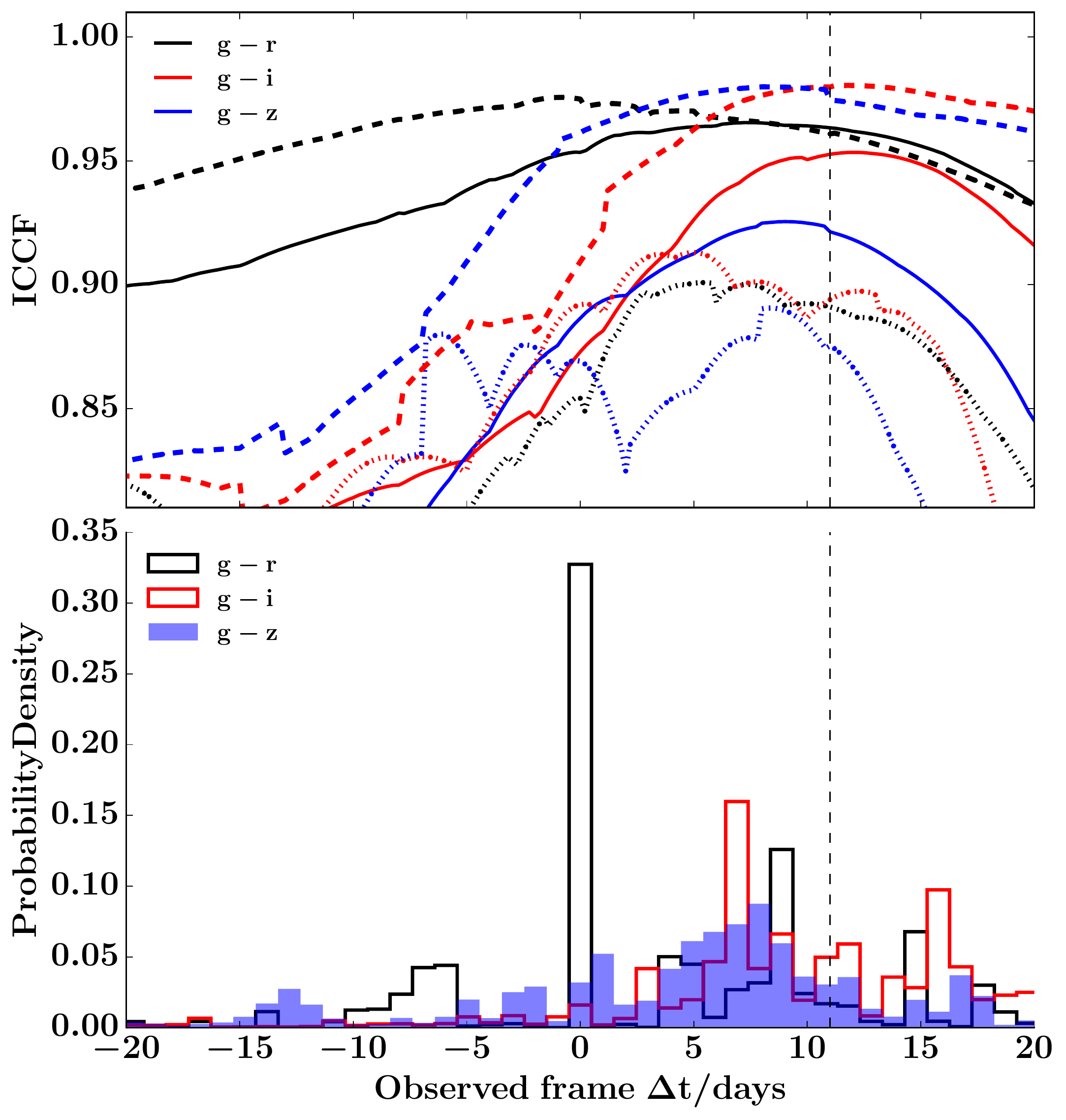}
\caption{Top: interpolated cross-correlation coefficient (ICCF) 
between the mock $g$ and $r$, $i$, $z-$band 
light-curves. The solid lines use all the data points,
while the dashed and dotted lines are based on
the first or second half of the data, to show the variations 
in the ICCF.
Bottom: probability density distributions of lags calculated 
based on the centroid of ICCF as described in Section \ref{sec:testcc}. 
This experiment is done for quasar J022020.02-034331.1 as 
in the bottom panel of Figure \ref{testbandlag}. From left to right, 
the mock $r$, $i$ and $z-$band light-curves 
are shifted by $2$, $5$ and $11$ days as indicated by the vertical dashed line. }
\label{testcclag}
\end{figure*}

\subsection{Test of the  Cross-Correlation Method}
\label{sec:testcc}

In order to assess how well the cross-correlation method is able to 
identify lags given the sampling of our light curves, we have
also tried this method following the procedure described in
\cite{Petersonetal2004} with the same mock light-curves as in the last
section.  We calculate the standard cross-correlation function (CCF)
with linear interpolation between data points for mock light-curves in
two different bands. The centroid of the CCF is calculated for points
with cross-correlation coefficient larger than $80\%$ of the peak
value of CCF.  We use 50000 independent realizations of the
light-curves to estimate the cross-correlation centroid distribution,
which corresponds to the probability distribution of the lags. An
updated ``flux randomization/random subset selection" (FR/RSS) method
\citep[][]{Petersonetal1998,Welsh1999} that accounts for the redundant
selections by reducing the flux uncertainties by the square root of
the number of multiple selections is used to estimate the centroid
distribution.  Figure \ref{testcclag} shows the results for the same
mock light-curves of quasar J022020.02-034331.1 as used in the bottom
panel of Figure \ref{testbandlag}.  Although the mock $r$, $i$ and 
$z-$band light-curves are shifted by $2$, $5$ and $11$ days respectively, the
cross-correlation method is unable to pick out the signals with the
cadences of our Pan-STARRS light-curves. 
Even if we only calculate CCF for the light curves in each season 
and average the results of all the seasons to avoid the large seasonable gaps, 
the cross-correlation method still cannot reproduce the input lags.
 In contrast, 
{\sf JAVELIN} succeeds at this test, as 
shown in Figure \ref{testbandlag}. Thus, 
we will only focus on the results calculated by {\sf JAVELIN}.

\subsection{Emission Lines} 

Our sample has also been observed spectroscopically
\citep[][]{Shenetal2015}, which allows us to quantify the
contamination of lines in our sample and study the relation between
the lags and various line equivalent widths (see section
\ref{sec:discussion}).

In our analysis of the inter-band lags, we only use one DRW model to
describe the light-curve in each band. If there are broad emission
lines contributing a significant fraction of the flux in each band,
they can affect the lags between the continuum radiation in different
bands we try to measure
\citep[][]{CheloucheZucker2013,Edelsonetal2015,Fausnaughetal2016},
because they have different lags with respect to the continuum
radiation.  Although we have chosen two special redshift ranges to
minimize the contamination, Figure \ref{filterspectrum} shows that
there are still some major lines in the bands, particularly blended
\ion{Fe}{2} and \ion{Mg}{2} in $g$ band and \ion{Mg}{2} and
\ion{H$\beta$}{0} in $r$ band, that can potentially contaminate the
lag signals.
 
In both the $g$ and $r$ bands, the ratio between the line and continuum
fluxes is always smaller than $10\%$ in our sample with a median value
$1\%$ in $g$ band and $3\%$ in $r$ band.  For the subsample \emph{cLD}
(see section \ref{sec:wholesample}), the ratio is always smaller than
$5.8\%$ with a median value $0.47\%$ in $g$ band and $3.2\%$ in $r$
band. Given the small ratio of line to continuum flux, in general the
broad lines cannot significantly affect the lags we detect \citep[][]{Fausnaughetal2016}.

\section{Results}
\label{sec:results}

\begin{figure*}[htp]
\centering
\includegraphics[width=0.48\hsize]{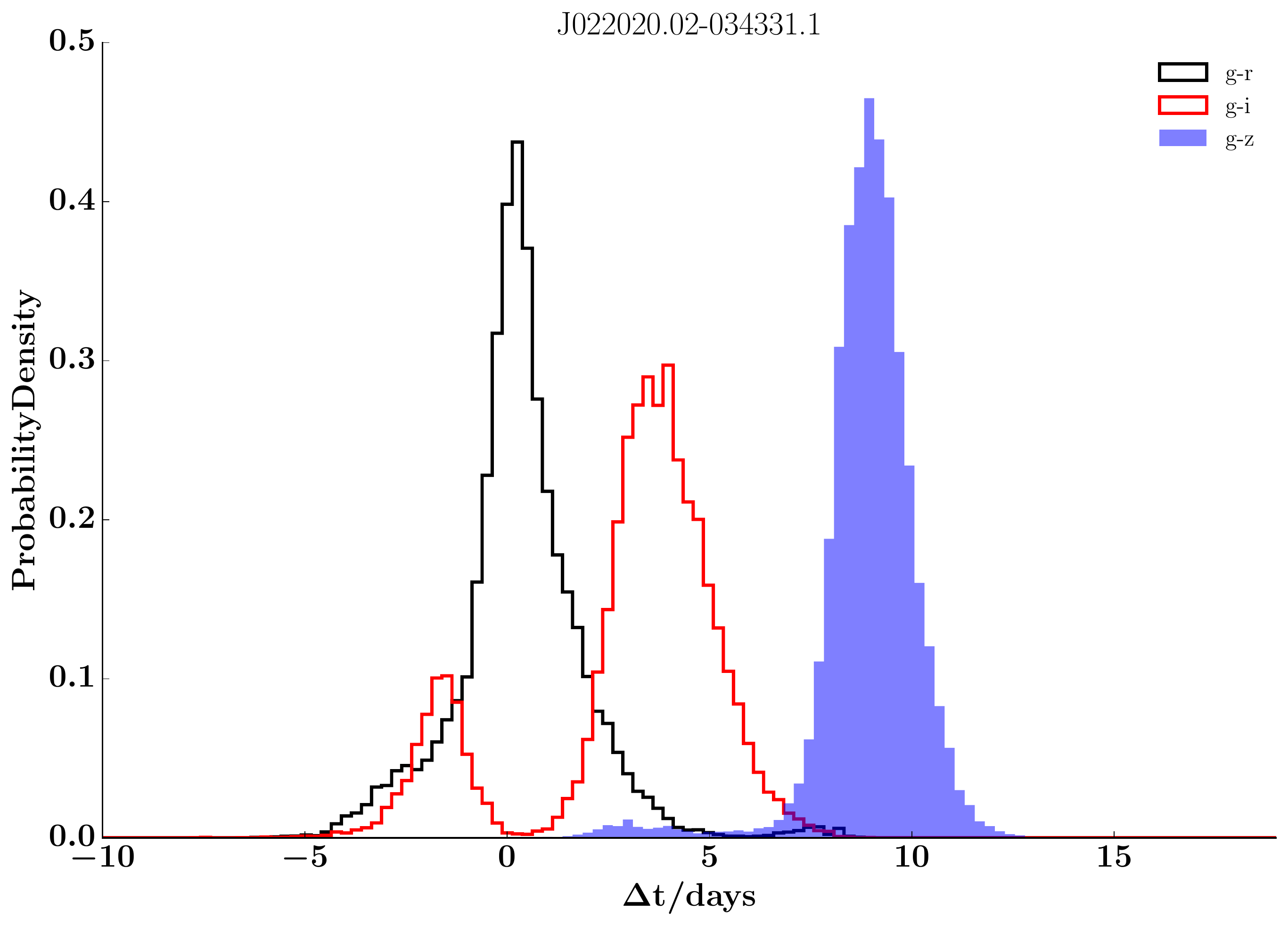}
\includegraphics[width=0.48\hsize]{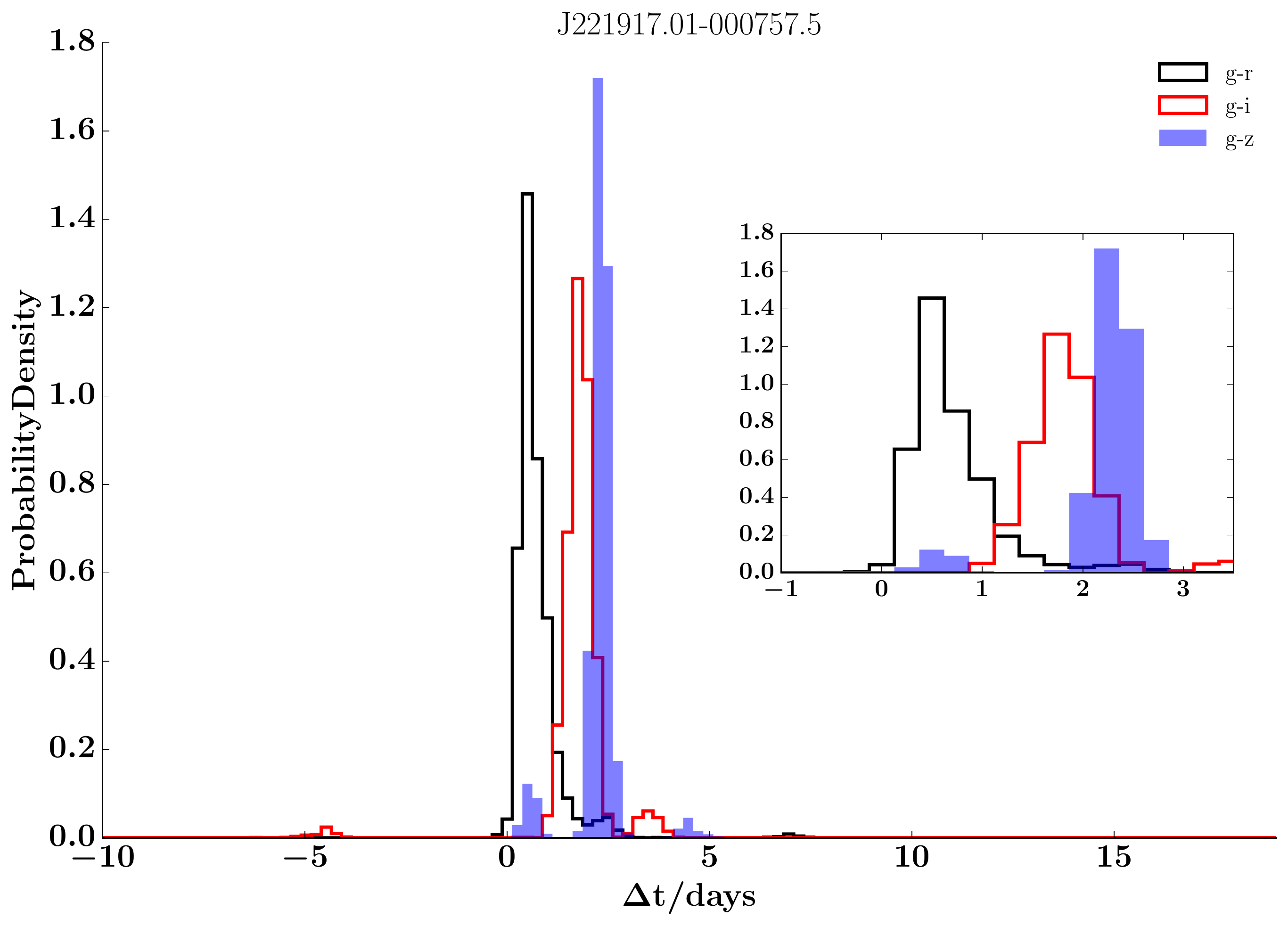}
\caption{Example histograms of the rest frame lags between $g$, $r$, $i$ 
and $z-$bands for the quasars J022020.02-034331.1 in MD01
and J221917.01-000757.5 in MD09. The small window in the right 
panel is the zoomed in plot between $-1$ and $3.5$ days. 
The two examples 
show cases with significantly detected lags. Histograms of 
rest frame lags for all quasars are available
online.}
\label{GoodExample}
\end{figure*}

\subsection{Lags for the Whole Sample}
\label{sec:wholesample}

\begin{figure}[htp]
\centering
\includegraphics[width=1\hsize]{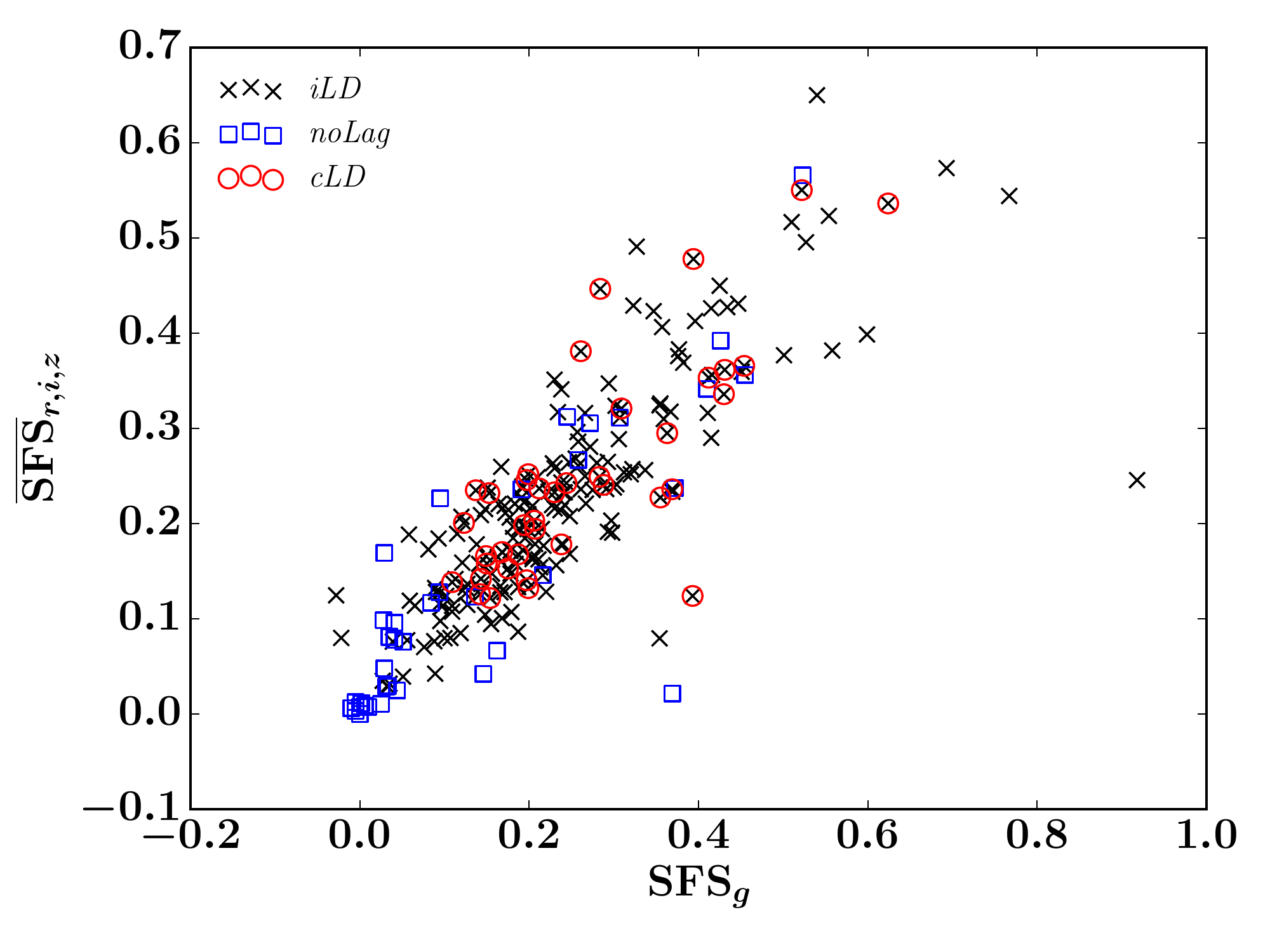}
\caption{Distributions of structure function slope (SFS) for the $g$ 
band light-curves and the mean SFS for light-curves in $r,i,z$ bands. 
The black crosses, red circles and blue squares are for samples 
\emph{iLD}, \emph{cLD} and \emph{noLag} respectively as explained 
in Table \ref{table:numbers}.}
\label{sfslope}
\end{figure}

\begin{figure}[h]
\centering
\includegraphics[width=1\hsize]{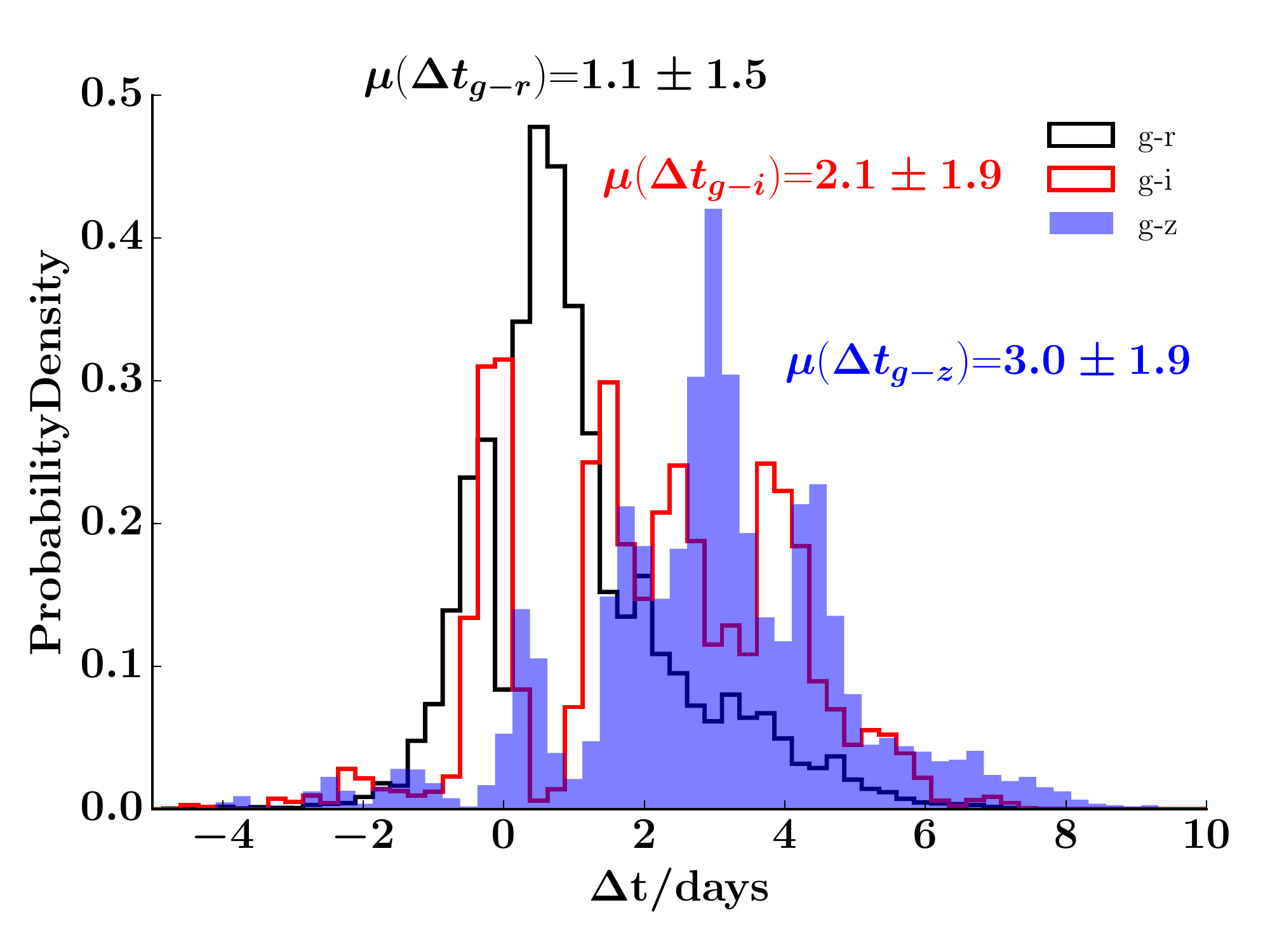}\\
\includegraphics[width=1\hsize]{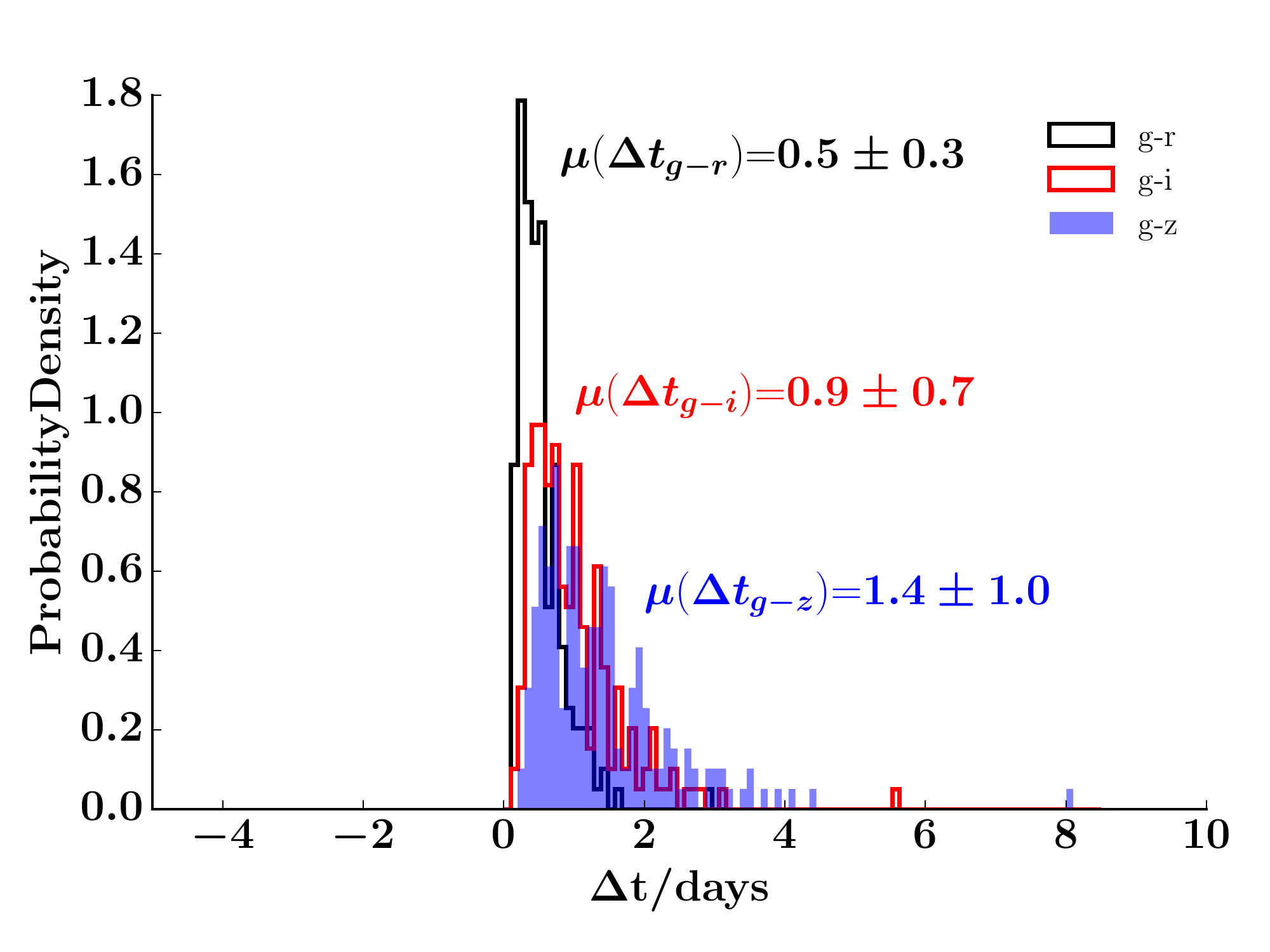}
\caption{Top: stacked histograms of lags between $g$ and $r,i,z$ bands for the
whole sample. The lags are measured in the rest frame of the
quasars. The probability density is normalized such that the total
area under each histogram is one. The averaged lags for each histogram
are labeled in the figure. Bottom: histograms of the theoretically
estimated lags between $g$ and $r,i,z$ bands based on the standard
thin disk model as described in Section \ref{sec:theory}, using the
estimated black hole mass and bolometric luminosity for each quasar.}
\label{LagAllSample}
\end{figure}

From our original sample of 240 quasars,  {\sf JAVELIN} is able to fit
DRW models and show probability density distributions of 
lag signals with gaussian shapes for 200
quasars (subsample \emph{iLD}). 
We have also done the Anderson-Darling test to 
make sure the fitting residuals do follow the Gaussian distributions.
However, not every quasar shows
strong and consistent variability across the four bands, which is
necessary for {\sf JAVELIN} to detect significant lags between the 
$g-$band and other bands.  Figure \ref{GoodExample} shows two examples of
the probability density distributions of the lags for quasars
J022020.02-034331.1 and J084536.18+453453.6 with significantly
detected signals. There is only a single dominant peak in each case,
which is very similar to the experiment shown in Figure
\ref{testbandlag}.  The lags between the $g-$band and other bands increase
with increasing wavelength.  For examples like this, we take the full
width at half maximum (FWHM), corresponding to 2.35$\sigma$ for Gaussian
distributions, as the uncertainty in the lag.  The lag is calculated as the
centroid of the distribution, which is the probability density
weighted mean lag in the region with probably density larger than half
of the maximum value.
 
 \begin{figure}[h]
\centering
\includegraphics[width=1\hsize]{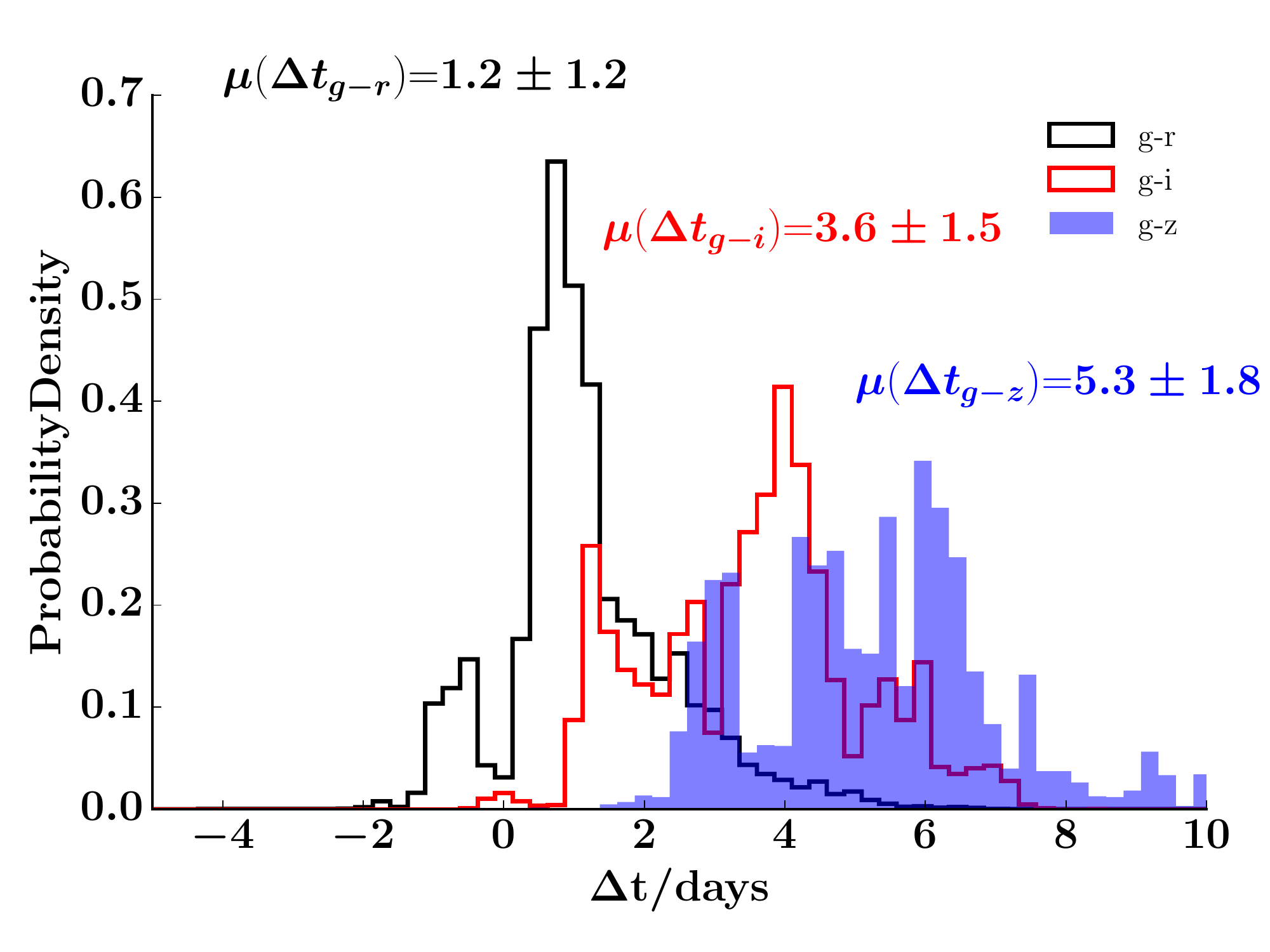}\\
\includegraphics[width=1\hsize]{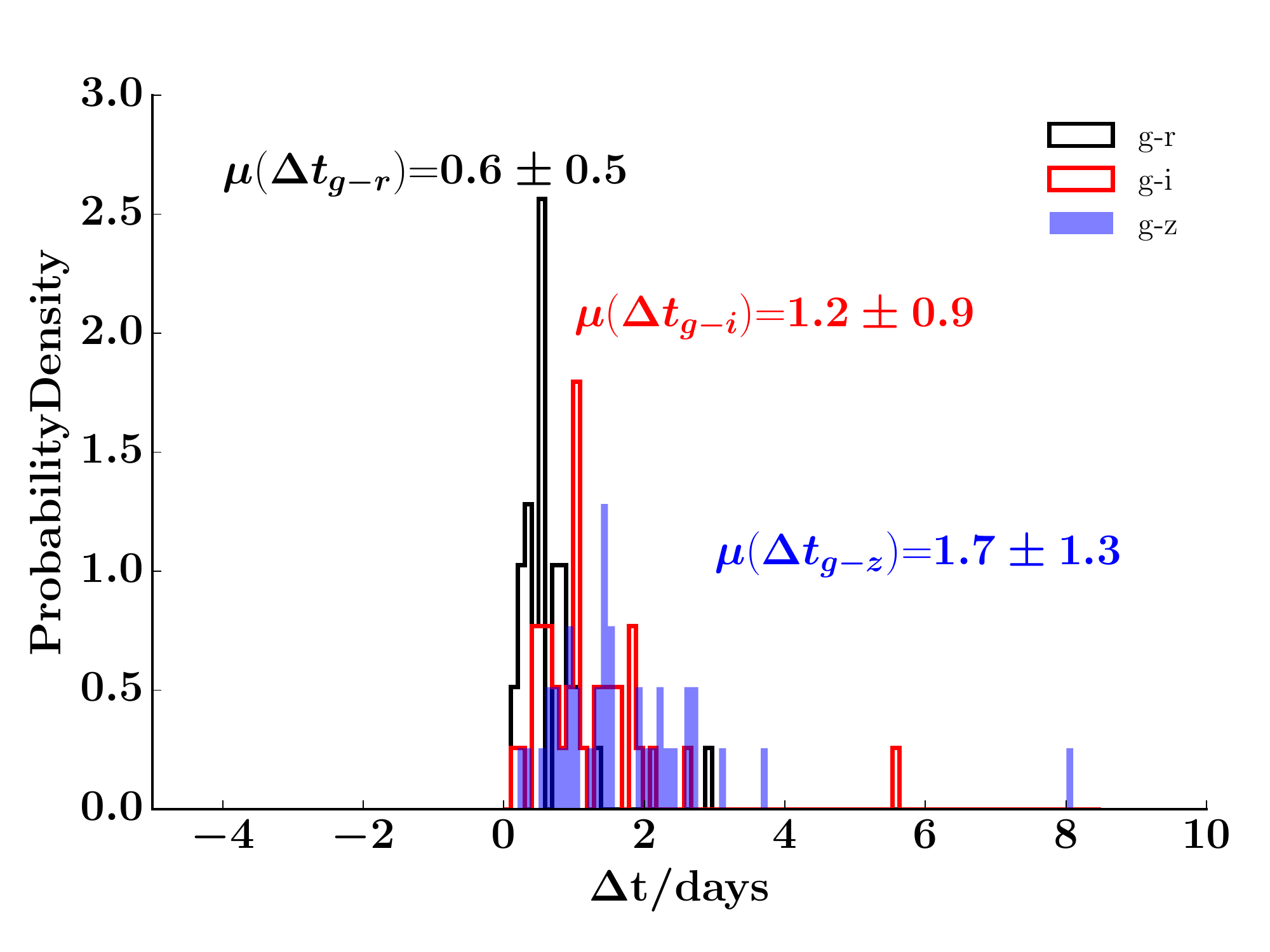}
\caption{Top: stacked histograms of lags between $g$ and $r,i,z-$bands 
for the selected sub-sample \emph{cLD}.  
Bottom: histograms of the theoretically estimated lags 
between $g$ and $r,i,z$ bands for this sub-sample.}
\label{LagSubSample}
\end{figure}

As quasars in each MDF have the same cadence, it is interesting to assess
whether the $40$ quasars (named subsample \emph{noLag}) for which
\emph{JAVELIN} cannot detect any lag signal have any property that is
significantly different from the others. The structure function
\citep[e.g.,][]{Choietal2014,MacLeodetal2010,Kozlowski2016}, which quantifies 
variability as a function of time scale, is a useful quantity
to show differences in variability properties across the
sample. We first calculate the structure functions for the $g,r,i,z-$band 
light curves for each quasar according to the procedure described
in \cite{Kozlowski2016}.  The structure function typically reaches a
maximum value around the damping time scale $\tau$ as in the DRW
model. For time differences smaller than $\tau$, the structure 
function is well-fit by 
a power law, the slope of which tells us how the variability amplitude
changes from short to long time scales. 

If the light curves have strong noise at time
scales smaller than the few day timescales in which we are interested,
the power law slope will be shallow and the DRW models will have difficulty
fitting the short time scale fluctuations and finding a lag.  If
the $r,i,z-$band light curves follow the $g-$band light-curves with a
fixed lag, we also expect them to have a similar structure function
slope (SFS). We calculate SFS$_g$, SFS$_r$, SFS$_i$ and SFS$_z$ for
subsamples \emph{iLD}, \emph{noLag} and \emph{cLD} (the last will be
defined below).  Figure \ref{sfslope} shows the distribution between
SFS$_g$ and the mean slope
$\overline{\rm{SFS}}_{r,i,z}\equiv\sqrt{({\rm SFS}^2_r+
  \text{SFS}^2_i+\text{SFS}^2_z)/3}$.  Most of the quasars in
\emph{noLag} are indeed located at the bottom left corner with small
SFS$_g$ and $\overline{\rm {SFS}}_{r,i,z}$, although they have similar
structure function normalizations.  This means that their
structure functions are flatter and their light-curves have more
variability/noise on short time scales. We have also checked that these quasars do
not have any special properties in terms of black hole mass,
luminosity, or normalized excess variance compared with the other
quasars.
 
For the other 200 quasars where \emph{JAVELIN} detects a lag, 
there are 102 quasars showing single dominant peaks between
the $g-r$, $g-i$ and $g-z$ bands while the rest have multiple peaks
distributed in a wide range from $-40$ days to $40$ days. The two
groups of quasars do not show any significant differences in the 
key parameters $L,\mbh,\sigma^2_{\rm rms}$ compared with each other.  
They also have similar ratios between the magnitude uncertainties and 
magnitudes. If we require
that lags increase from $g-r$, $g-i$ to $g-z$ bands, as they would if
the lags are caused by reprocessing of radiation from the center, we
find 39 quasars in our sample (named subsample \emph{cLD}).  The physical
properties of this subsample are shown by the open red squares in
Figure \ref{SampleProperty}.  They span the whole parameter
space of the original sample, and again do not show any systematic
difference compared with \emph{iLD} in their luminosity, black hole mass
or normalized excess variance distributions.  The median
luminosity of this subsample is $5.4\times 10^{45}$ erg s$^{-1}$, which is
consistent with the median luminosity of the whole sample. 

For the other 63 quasars that have single dominant peaks but do not 
show a progression in time lag for the redder bands, 
23 quasars have both $g-i$ and $g-z$ lags smaller
than $g-r$ lags, while 20 quasars only have $g-i$ lags smaller than
$g-r$ lag and $11$ quasars only have $g-z$ lags smaller than $g-r$
lags. There are also 9 quasars showing both $g-i$ and $g-z$ lags
larger than $g-r$ lags but with $g-z$ lags smaller than $g-i$ lags.
The numbers of quasars for the initial sample and subsamples
\emph{noLag}, \emph{iLD}, \emph{cLD} are summarized in Table
\ref{table:numbers}.



\subsection{The Stacked Signals}
\label{sec:stacking}

In order to see the properties of lags for the whole sample, we stack
the signals in the following way.  We divide the lag interval from
$-20$ to $20$ days into $160$ bins with bin width $0.25$ day.  For
each quasar, we count the number of MCMC samples yielding the rest frame lag in
each bin. We then calculate the median value of the MCMC counts for
the whole sample for each bin, which gives the probability
distribution of the stacked lags for the whole sample. We repeat the
process for the $g-r$, $g-i$ and $g-z$ lags. Because each quasar has
the same number of total MCMC trials, all quasars are given the 
same weight, thus building up the
distribution of most likely lag for the sample as a whole. Probability
density distributions of the rest frame lags stacked in this way are shown in the
top panel of Figure \ref{LagAllSample}.

The probability density weighted mean lags for the $g-r$, $g-i$ and
$g-z$ bands are $1.1\pm 1.5$, $2.1\pm1.9$ and $3.0\pm1.9$ days in the
rest frame with the quoted uncertainties being the standard deviation  
and most of the lag signals being positive. The mean lags 
increase from $g-r$ to $g-z$ bands, although they are still consistent within 
one standard deviation. Differences between the mean
$g-r$, $g-i$ and $g-z$ lags should be proportional to the distances between
the radii where $g$, $r$, $i$ and $z$ photons are expected to be
emitted, which will be discussed in Section \ref{sec:theory} in
detail.  The stacked signals have broad distributions and multiple peaks, particularly
for the $g-i$ and $g-z$ lags.  This is partially because quasars in our
sample have a wide range of luminosities and black hole masses, as shown
in Figure \ref{SampleProperty}.  Another reason is that apart from the
\emph{cLD} subsample, individual quasars do not have single
peaked and well ordered lags, which will contribute to the noise in
the stacked profile. In order to see whether the stacked signals are
dominated by the subsample \emph{cLD} or not, we have repeated the
same stacking experiment excluding the quasars in the \emph{cLD}
subsample. We find very similar stacked signals as shown in Figure
\ref{LagAllSample}, which suggests that the results represent
properties of the whole sample.

\begin{table}
\centering
\caption{Rest Frame Lags for the subsample \emph{cLD}}
 \label{table:lag}
\scalebox{1.0}{
\begin{tabular}{cccc}
\hline \hline
                      &  $\Delta t_{g-r}$ & $\Delta t_{g-i}$ & $\Delta t_{g-z}$ \\
SDSS Name  &     days    &     days   &     days     \\
 \hline
J022659.82-035015.0 & $-0.13\pm1.75$ & $1.50\pm1.00$ & $2.98\pm2.00$ \\ 
J022144.75-033138.8 & $4.27\pm1.00$ & $7.34\pm0.75$ & $13.75\pm0.25$ \\ 
J022020.02-034331.1 & $0.15\pm1.25$ & $3.81\pm2.25$ & $9.08\pm1.75$ \\ 
J022340.29-042852.4 & $-0.11\pm3.25$ & $1.74\pm1.50$ & $1.73\pm3.50$ \\ 
J084536.18+453453.6 & $0.40\pm0.75$ & $6.78\pm1.00$ & $11.95\pm3.25$ \\ 
J084512.99+445208.9 & $2.11\pm1.75$ & $4.59\pm2.25$ & $4.46\pm1.50$ \\ 
J083841.70+430519.0 & $1.09\pm0.25$ & $1.59\pm0.25$ & $5.91\pm0.25$ \\ 
J084610.76+452153.1 & $2.18\pm0.75$ & $5.38\pm0.25$ & $5.34\pm0.75$ \\ 
J084341.41+444023.3 & $0.10\pm1.75$ & $1.12\pm0.75$ & $10.05\pm1.50$ \\ 
J083756.22+431713.4 & $-0.83\pm1.75$ & $5.81\pm1.00$ & $8.97\pm1.00$ \\ 
J083836.14+435053.3 & $3.49\pm2.50$ & $5.87\pm0.75$ & $12.63\pm0.75$ \\ 
J083425.01+442658.2 & $-0.02\pm1.00$ & $1.03\pm0.50$ & $2.22\pm1.50$ \\ 
J084517.64+441004.9 & $0.75\pm0.25$ & $3.25\pm0.25$ & $4.65\pm0.25$ \\ 
J095701.58+023857.3 & $5.30\pm0.50$ & $6.40\pm0.25$ & $9.00\pm0.25$ \\ 
J100029.15+010144.8 & $-0.25\pm0.25$ & $1.50\pm0.25$ & $2.00\pm0.25$ \\ 
J100421.01+013647.3 & $2.36\pm2.75$ & $4.60\pm0.25$ & $12.10\pm0.75$ \\ 
J100327.67+015742.4 & $4.15\pm0.75$ & $3.99\pm0.50$ & $8.89\pm0.25$ \\ 
J100025.24+015852.0 & $-0.00\pm0.50$ & $4.00\pm0.25$ & $7.84\pm0.25$ \\ 
J122549.28+472343.7 & $0.61\pm0.25$ & $2.38\pm0.25$ & $2.65\pm0.25$ \\ 
J142336.76+523932.8 & $0.07\pm0.75$ & $4.01\pm1.00$ & $11.66\pm0.75$ \\ 
J141104.86+520516.8 & $5.51\pm1.00$ & $6.73\pm0.50$ & $14.16\pm0.75$ \\ 
J141018.04+523446.0 & $2.96\pm1.50$ & $3.97\pm0.50$ & $7.38\pm0.25$ \\ 
J140739.16+525850.7 & $0.75\pm0.25$ & $3.51\pm0.50$ & $10.50\pm0.25$ \\ 
J141147.59+523414.5 & $-1.50\pm0.50$ & $0.88\pm1.25$ & $4.37\pm0.25$ \\ 
J141539.59+523727.9 & $-0.62\pm0.25$ & $1.75\pm0.25$ & $2.75\pm0.25$ \\ 
J142008.27+521646.9 & $3.35\pm0.25$ & $7.00\pm0.25$ & $6.86\pm0.25$ \\ 
J141138.06+534957.7 & $0.75\pm0.50$ & $3.21\pm1.00$ & $3.54\pm0.50$ \\ 
J141811.34+533808.5 & $0.23\pm3.00$ & $2.01\pm2.00$ & $3.16\pm3.75$ \\ 
J141358.90+542705.9 & $-0.20\pm1.00$ & $2.89\pm3.25$ & $3.61\pm1.75$ \\ 
J141856.19+535844.9 & $2.09\pm0.25$ & $9.25\pm0.25$ & $14.34\pm0.25$ \\ 
J142106.26+534406.9 & $-0.75\pm0.50$ & $7.00\pm0.25$ & $11.89\pm0.25$ \\ 
J221504.35+010935.2 & $2.61\pm5.25$ & $5.05\pm6.25$ & $10.86\pm1.75$ \\ 
J221434.82+001923.9 & $-0.25\pm1.50$ & $1.01\pm0.50$ & $3.27\pm2.50$ \\ 
J221447.75-002032.7 & $3.35\pm0.25$ & $4.73\pm1.00$ & $6.36\pm0.75$ \\ 
J221917.01-000757.5 & $0.59\pm0.25$ & $1.78\pm0.50$ & $2.36\pm0.25$ \\ 
J222228.39+002640.6 & $1.17\pm2.00$ & $5.65\pm0.75$ & $8.78\pm1.00$ \\ 
J232826.57+010207.8 & $2.26\pm0.50$ & $5.24\pm0.50$ & $6.12\pm0.75$ \\ 
J232907.12+003416.6 & $1.70\pm1.50$ & $4.25\pm0.50$ & $6.51\pm1.00$ \\ 
J233201.42-005655.2 & $3.50\pm1.50$ & $5.76\pm0.50$ & $6.23\pm0.50$ \\ 
 \hline
\end{tabular}}
\begin{center}
\begin{tablenotes}
\item Note: properties of these lags are discussed in Figure \ref{lagSubLumBH}, \ref{laglag} and Figure \ref{negativelagsigma}.
\end{tablenotes}
\end{center}
\end{table}

\subsection{Dependence of Lags on Luminosity}

\begin{figure}[htp]
\centering
\includegraphics[width=1\hsize]{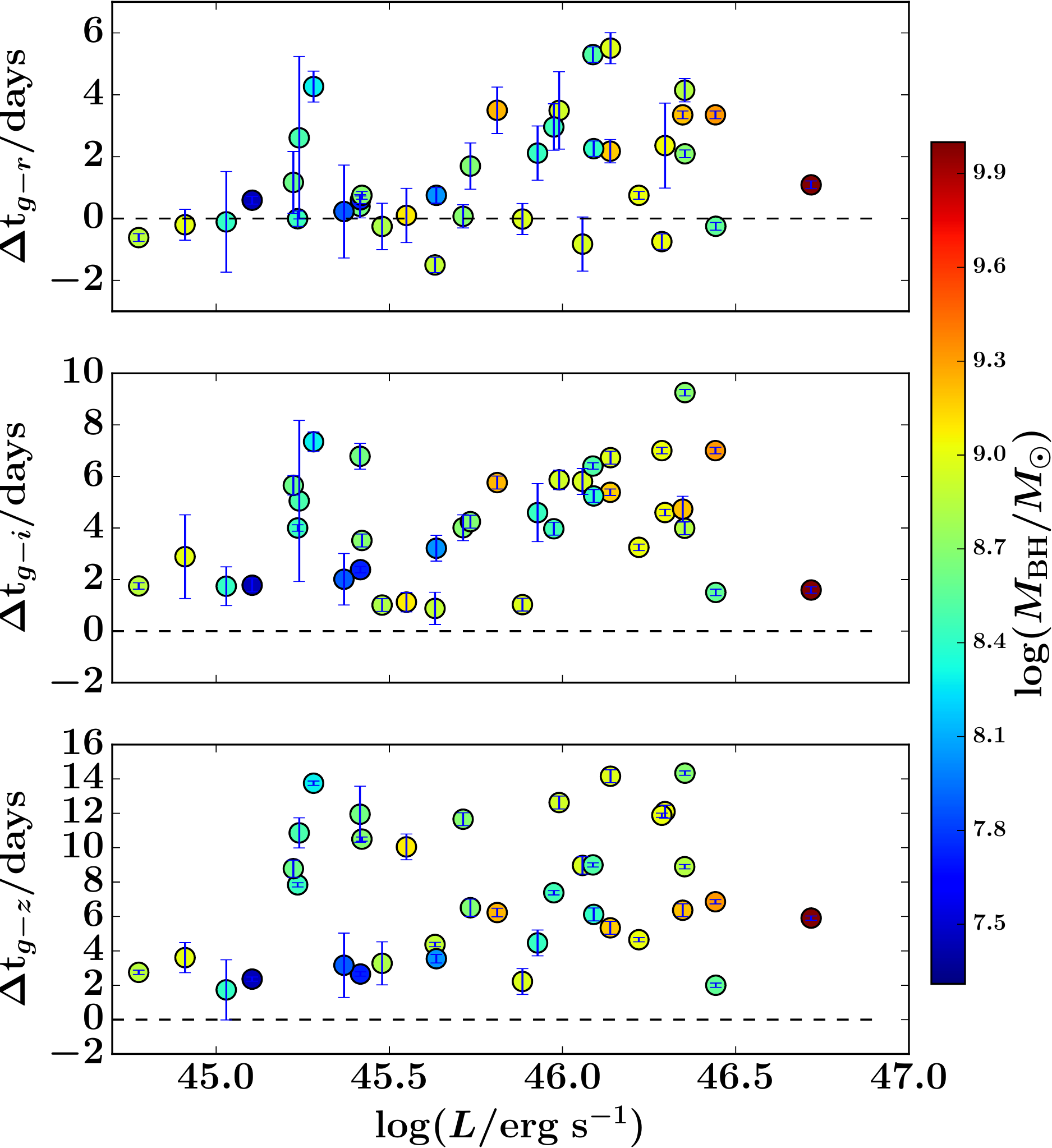}
\caption{
Distributions of lag with luminosity for the $39$ \emph{cLD} 
quasars with significant, consistent detections. 
Each data point is color-coded by the estimated black hole mass. 
From top to bottom, the three panels 
are rest frame lags between $g-r$, $g-i$ and $g-z$ bands respectively. 
The dashed black lines in each panel indicate 0 lags. }
\label{lagSubLumBH}
\end{figure}

\begin{figure*}[htp]
\centering
\includegraphics[width=0.48\hsize]{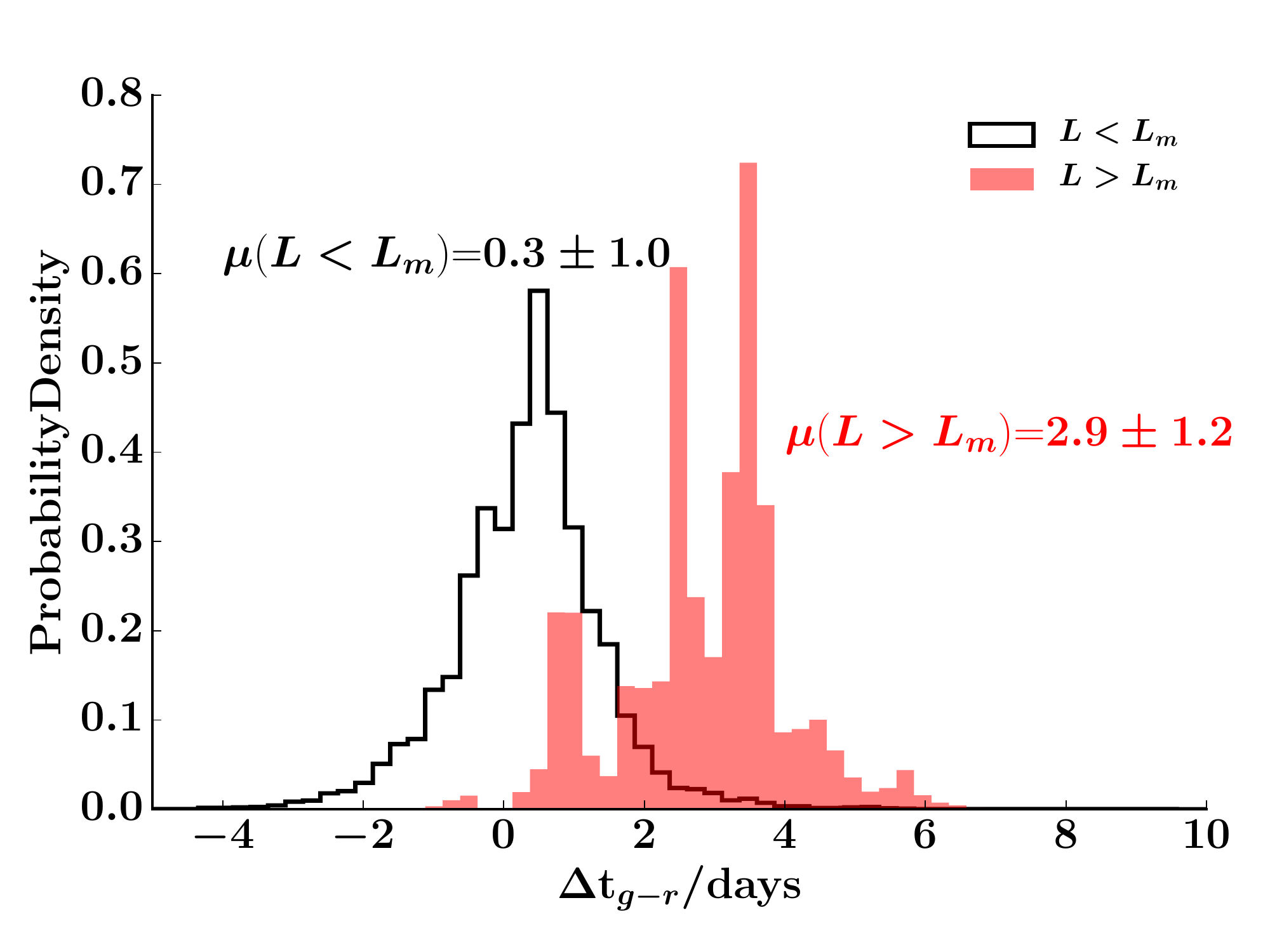}
\includegraphics[width=0.48\hsize]{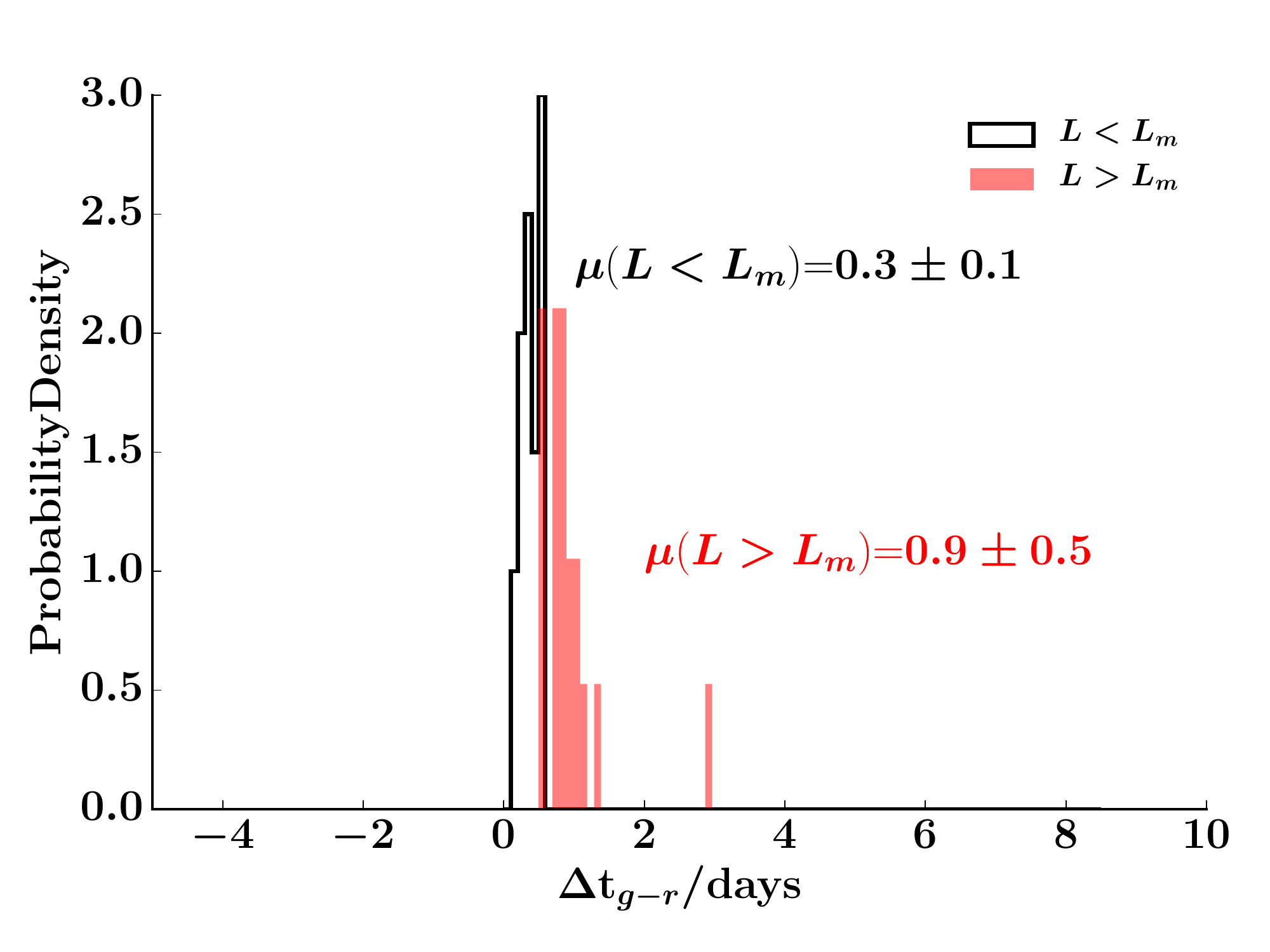}
\includegraphics[width=0.48\hsize]{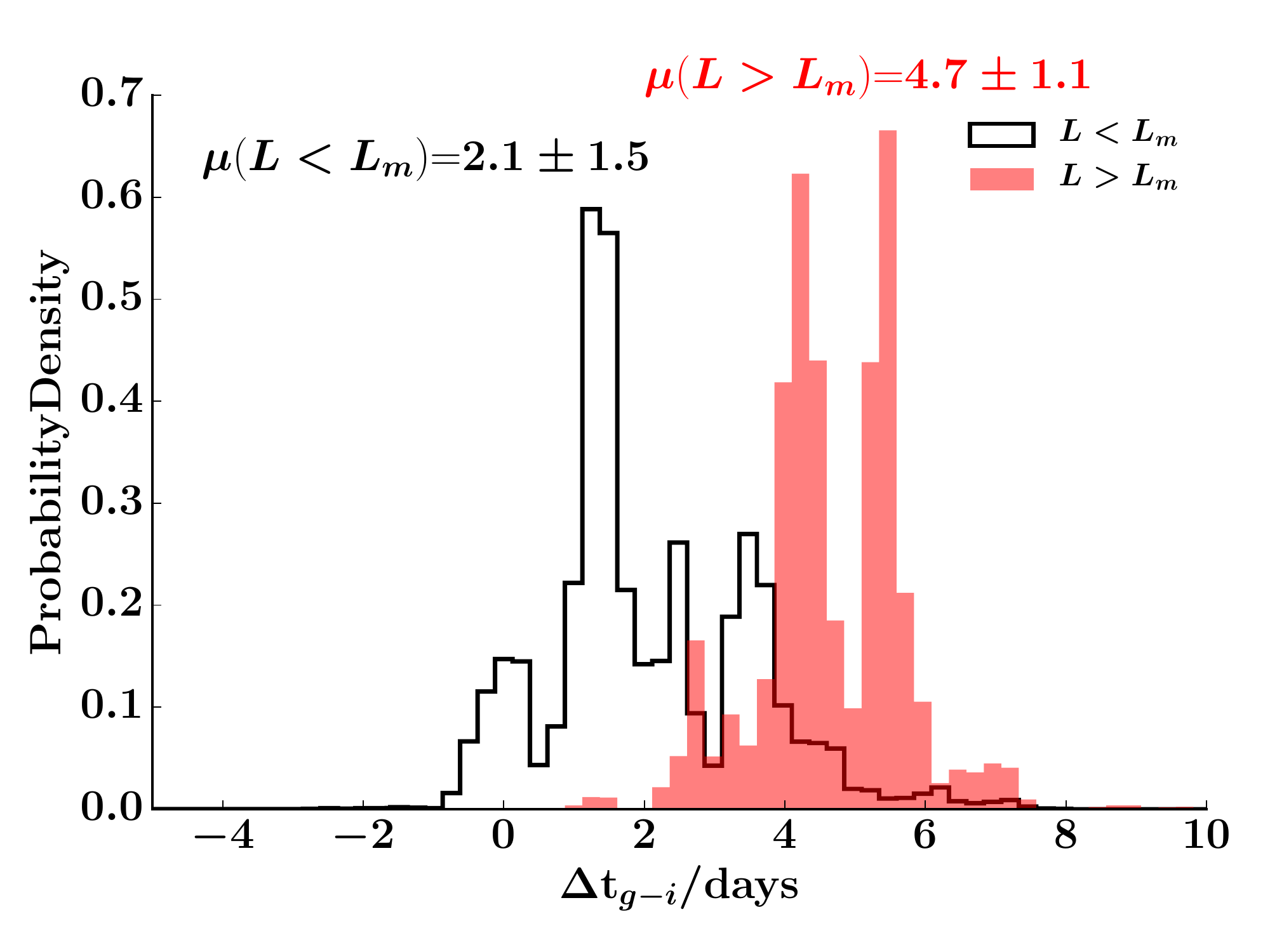}
\includegraphics[width=0.48\hsize]{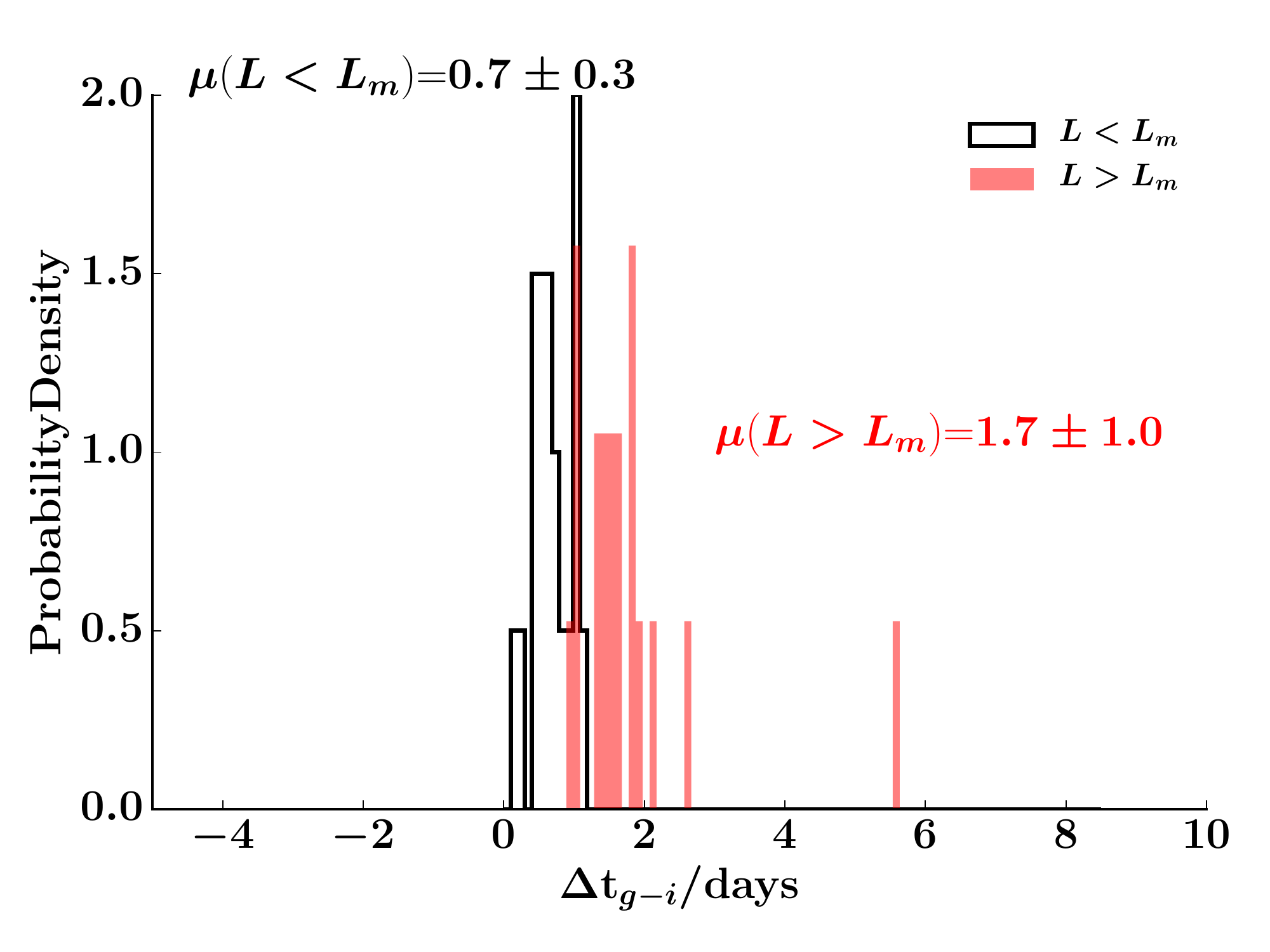}
\includegraphics[width=0.48\hsize]{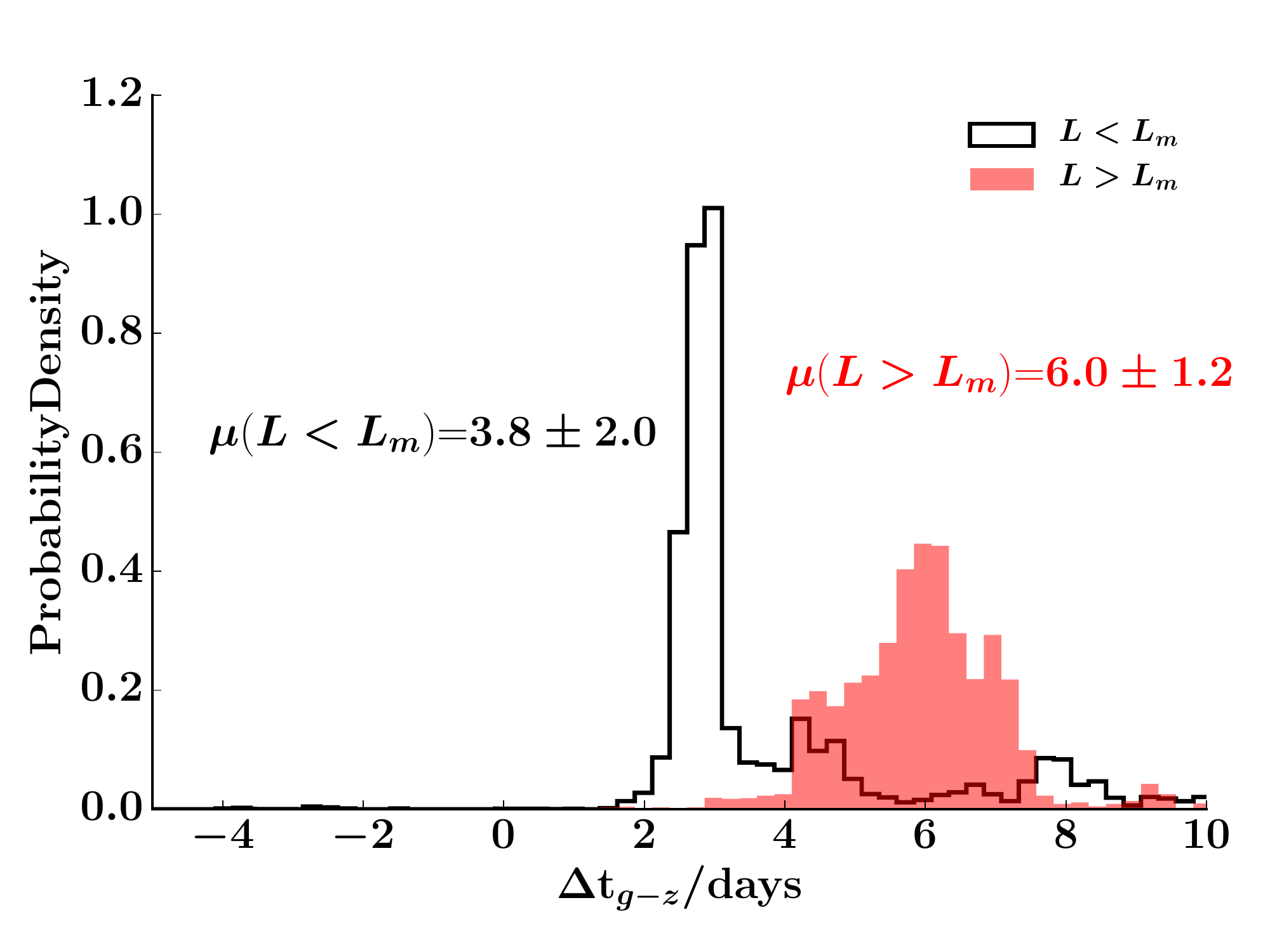}
\includegraphics[width=0.48\hsize]{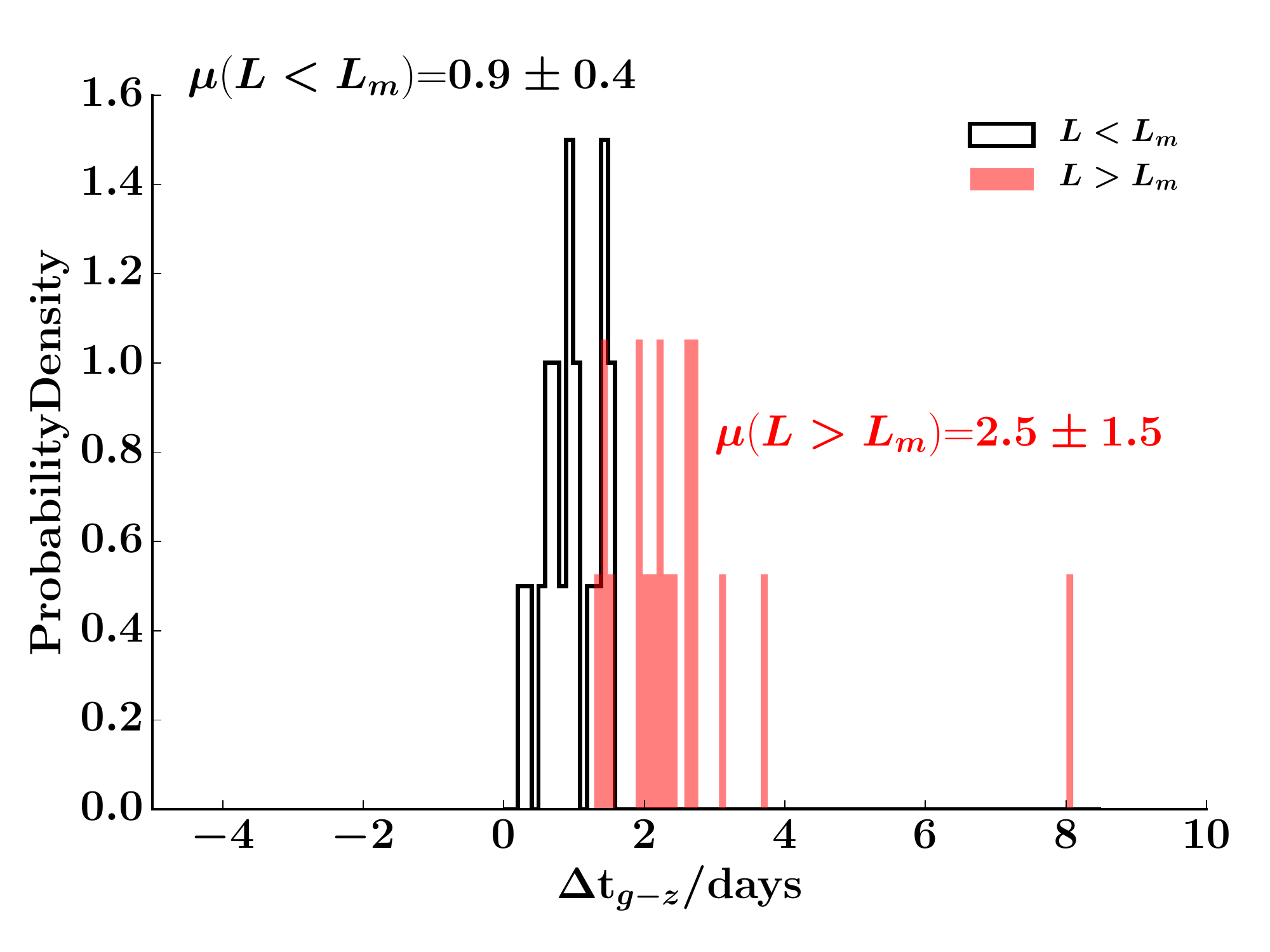}
\caption{Left: stacked histograms of the lags between the 
$g-$band and the other bands, when 
the 39 \emph{cLD} quasars with significant, consistent lag detections are divided 
into two sub-samples based on the median luminosity $L_m$. 
The solid black line is the sub-sample with $L<L_m$, while the dashed red line 
is the sub-sample with $L>L_m$. 
Right: histograms of the theoretically estimated lags between the $g$ 
and other bands for the two subsamples divided by luminosity in the same way. 
From top to bottom, the three rows are for lags between $g-r$, $g-i$ and  
$g-z$ bands respectively. }
\label{lagSubLum}
\end{figure*}

If the detected lags are related to the light-travel time across
different radii of the accretion disks, they should vary with the
luminosity and black hole mass of the quasars, since these determine
the sizes of accretion disks. The increase of the interband lags with
increasing luminosity has also been noticed by \cite{Sergeevetal2005}
based on observations of 14 AGNs.  We check for luminosity dependence
by dividing the whole sample into two subsamples, one with luminosity
smaller than the median value $3.7\times 10^{45}$ erg s$^{-1}$ and the other
one with luminosity larger than this value. We stack the quasars in
each luminosity bin as described in section
\ref{sec:stacking}.  The mean lags between the $g-$band and the $r$, $i$
and $z$ bands for the lower luminosity subsample are $0.5\pm1.3$,
$0.7\pm1.8$ and $2.1\pm2.8$ days respectively, while the corresponding
mean lags for the higher luminosity subsample are $2.0\pm1.6$,
$3.8\pm1.3$ and $3.3\pm1.4$ days. The mean lags do increase with
luminosity as expected, although the uncertainty is large for the
stacked signals of the whole sample. The p values of  Kolmogorov-Smirnov (KS) test for the 
null hypothesis that the subsamples with high and low luminosities are 
from the same distributions for $g-r$, $g-i$ and $g-z$ lags are $6\times 10^{-10}$, 
$2\times 10^{-25}$ and $2\times 10^{-4}$ respectively, which supports the conclusion 
that the two subsamples are significantly different.

\subsection{Lags for the \emph{cLD} Subsample }
Lags for the 39 \emph{cLD} quasars with significant, well
ordered lag detections are summarized in Table \ref{table:lag}.  
We repeat the same stacking process for this subsample to look
for the mean lags between different bands, which is shown in the top
panel of Figure \ref{LagSubSample}.  The probability density weighted
mean $g-r$, $g-i$ and $g-z$ lags for the subsample are $1.2\pm1.2$,
$3.6\pm1.5$ and $5.3\pm1.8$ days. The lags between the $g$ and $r$
bands are almost the same as the stacked lags for the whole sample,
while lags between $g$ and $i, z$ bands become larger.  The
probability density of negative lags for $g-i$ and $g-z$ bands is
almost zero for this subsample.  This subsample will be the focus of
the analysis in the following sections, mainly because only they can
be compared with simple reprocessing models to constrain the physics of
accretion disks individually.

Distributions of the lags as a function of luminosity and black hole mass
for the 39 \emph{cLD} quasars are shown in Figure
\ref{lagSubLumBH}. There is a weak trend whereby the lags increase
with increasing luminosity, although the scatter is large.  The
Pearson correlation coefficients between luminosity and $g-r$, $g-i$,
$g-z$ lags are $0.41$, $0.35$ and $0.24$ with corresponding $p$ values 
$0.0088$, $0.028$, $0.14$, while the 
Spearman correlation coefficients between luminosity and the three lags 
are $0.39$, $0.33$ and $0.24$
respectively with corresponding $p$ values $0.014$, $0.042$ and $0.14$. 
Here the $p$ values
represent the probabilities that  luminosity does not correlate with the lags. 
Least-squares fits to the lags and luminosities in 
log-log space (with negative lags excluded) give $\Delta t_{g-r}\propto
L^{0.55\pm0.37}$, $\Delta t_{g-i}\propto L^{0.16\pm0.16}$ and 
$\Delta t_{g-z}\propto L^{0.14\pm0.15}$. 
These fits are 
consistent with \cite{Sergeevetal2005} for the $g-r$
lags but show weaker dependence for $g-i$ and $g-z$ lags.

To quantify the trend with luminosity, we divide the 39 quasars into two subsamples
based on the median luminosity and repeat the stacking experiment as
before for each subsample.  The median luminosities for the two
subsamples are $2.11\times 10^{45}$ erg s$^{-1}$ and $1.38\times 10^{46}$
erg s$^{-1}$.  Probability density profiles of the lags between the $g-$band and
the $r,i,z-$bands for the two subsamples are shown in the left panel of
Figure \ref{lagSubLum}. For the low- and high- luminosity subsamples,
the mean lags between the $g$ and $r,i,z-$bands are $0.3\pm 1.1$, $2.1\pm
1.5$, $3.8\pm 2.0$ days and $2.8\pm 1.2$, $4.7\pm 1.1$ and $6.0\pm
1.2$ days, respectively, where the error bars represent the standard
deviation of the lags in the stacked probability density
distributions. {
The p values of KS test for the null hypothesis that the subsamples 
are from the same distributions for $g-r$, $g-i$ and $g-z$ lags are 
$4\times 10^{-6}$, $4\times 10^{-6}$ and $2\times 10^{-5}$ respectively.
}
This clearly shows that the averaged detected lags of
the high-luminosity quasars are significantly larger than lags
detected for the low-luminosity quasars.

\subsection{Comparison with the Standard Thin Disk Model}
\label{sec:theory}

In the standard thin disk model \citep[][]{ShakuraSunyaev1973}, the
effective temperature at each radius is determined by the local
dissipation rate and it changes with radius $R$ as $R^{-3/4}$ for a
fixed black hole mass and accretion rate.  If irradiation from the
inner region of the disk contributes significantly to the local
heating rate, the temperature profile may change. 

Following \cite{Fausnaughetal2016}, we assume the effective
temperature $T$ changes with radius $R$ as
\begin{eqnarray}
T(R)=\left(f_i\frac{3G\mbh\dot{M}}{8\pi\sigma R^3}\right)^{1/4},
\label{eqn:tr}
\end{eqnarray}
where $f_i$ is a factor that accounts for the irradiation from an
X-ray/UV source near the black hole by changing the normalization of
the temperature profile, $\sigma$ is the Stefan-Boltzmann constant,
$G$ is the gravitational constant, and $\dot{M}$ is the mass accretion
rate. The photon wavelength $\lambda$ is related to the characteristic
temperature $T$ as $T=hc/\left(Xk_{\rm B}\lambda\right)$, where $h$
and $k_{\rm B}$ are the Planck and Boltzmann constants, $c$ is the
speed of light, and $X=2.49$ \citep[][]{Fausnaughetal2016} is a factor
to account for emission at this wavelength from a range of
temperatures.  The central wavelengths of Pan-STARRS $g,r,i,z$ bands
are $4750$, $6250$, $7550$ and $8700$\AA, which are converted to rest-frame 
wavelengths using the redshift of each quasar. Based on these
relations, the median values of the radii in the accretion disks
corresponding to these wavelengths for our sample are $17.8r_s$,
$25.7r_s$, $33.0r_s$ and $39.9r_s$ respectively, where $r_s\equiv
2G\mbh/c^2$ is the Schwarzschild radius. Notice that the distances
between neighboring bands are roughly equal in this simple model.  The
light travel time across two different radii where photons with
wavelengths $\lambda_g$ and $\lambda_x$ are emitted are
\begin{eqnarray}
\Delta t_{g-x}=\left(X\frac{k_{\rm B}\lambda_g}{hc}\right)^{4/3}\left(f_i\frac{3G\mbh\dot{M}}{8\pi\sigma}\right)^{1/3}\nonumber\\
\left[\left(\frac{\lambda_x}{\lambda_g}\right)^{4/3}-1\right].
\label{eqn:lags}
\end{eqnarray}

For each quasar, we calculate the mass accretion rate based on the
bolometric luminosity and estimated black hole mass, assuming a
radiative efficiency of $10\%$. Then we stack the theoretically
calculated lags with $f_i=1$ for simplicity. (See discussion section
\ref{sec:discussion}.) for the whole sample and compare with our
detected lags, which are shown in the bottom panel of Figure
\ref{LagAllSample}. The same calculation is also done for the $39$
\emph{cLD} quasars and the stacked profiles are shown in
the bottom panel of Figure \ref{LagSubSample}. It is clear that the
theoretically calculated lags are
always smaller than the detected lags by a factor of $2$ for 
$g-r$ lags and a factor of $3$ for $g-i$ and $g-z$ lags. 
We also do the same calculations for the luminosity subsamples, which are
shown in the right panel of Figure \ref{lagSubLum}.  The theoretically
calculated lags indeed increase with increasing luminosity in our
sample. But they are always smaller than the detected lags, except for
the $g-r$ lags in the low luminosity subsample, where the lags are not
well constrained.

\begin{figure}[htp]
\centering
\includegraphics[width=1\hsize]{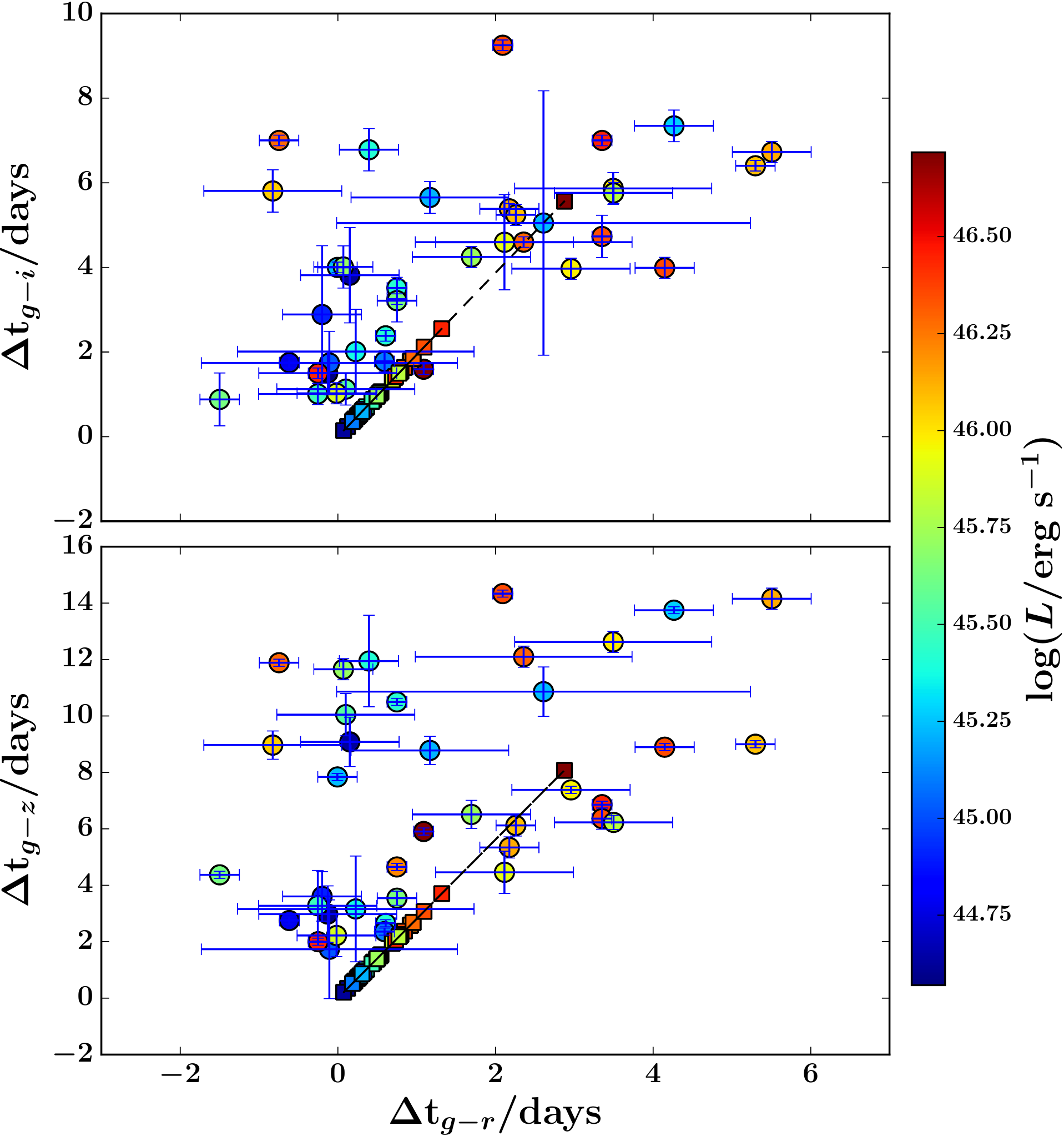}
\caption{Distributions of $g-r$ lags for the subsample \emph{cLD} 
with respect to $g-i$ (top panel) and $g-z$ (bottom panel) lags. 
Each data point is color coded with the bolometric luminosity. 
The filled squares connected by the dashed lines 
are theoretically calculated lags according to equation \ref{eqn:lags}.}
\label{laglag}
\end{figure}

Temperature profiles of the disk in the radial range $\sim 18r_s-40r_s$ 
can also be constrained by comparing 
$\Delta t_{g-i},\Delta t_{g-z}$ with $\Delta t_{g-r}$. According to Equation 
\ref{eqn:lags}, the three lags should be linearly proportional to each other as 
\begin{eqnarray}
\Delta t_{g-i}&=&\frac{\left(\lambda_i/\lambda_g\right)^{4/3}-1}{\left(\lambda_r/\lambda_g\right)^{4/3}-1}\Delta t_{g-r},\nonumber\\
\Delta t_{g-z}&=&\frac{\left(\lambda_z/\lambda_g\right)^{4/3}-1}{\left(\lambda_r/\lambda_g\right)^{4/3}-1}\Delta t_{g-r}.
\end{eqnarray}
For the fixed wavelength ratios between $r,i,z$ band and $g$ band, the slopes are only determined by the 
radial profile of effective temperature, which is $3/4$ in the standard thin disk model.   
The slopes are independent of black hole mass, luminosity, accretion rate and redshift, which only determine 
the actual values of the lags. 
The relations between the lags for the subsample \emph{cLD} and the theoretically calculated lags 
are shown in Figure \ref{laglag}. Compared with the actual
lags, the theoretically predicted relation falls
systematically on one side of the data.  
The best fitted
relations between bands are $\Delta t_{g-i}\propto \Delta
t_{g-r}^{0.25\pm 0.13}$ and $\Delta t_{g-z}\propto \Delta t_{g-r}^{0.08\pm0.17}$, 
where the error bars are for $95\%$ confidence level. 
This is significantly flatter compared with the theoretically expected linear
relation. If we force a linear relation between the lags, the best fittings are 
$\Delta t_{g-i}=(0.72\pm0.28)\Delta t_{g-r}+(3.08\pm0.71)$ and $\Delta t_{g-z}=(0.98\pm0.64)\Delta t_{g-r}+(5.86\pm1.42)$, 
where the theoretical slopes should be $1.94$ and $2.81$ respectively.

The differences in the correlations point either to a different 
temperature profile than we assume here, or a different relationship 
between lags and $\mbh$ and $\dot{M}$, particularly as the $g-r$ lag gets larger
(corresponding to high luminosity quasars). 
 A flatter
temperature profile and larger apparent disk size are both natural
consequences of the model proposed by \cite{Lawrence2012}, where the
radiation is reprocessed by some cold, thick clouds (see also
\citealt{GardnerDone2016}). However, alternative pictures, such as
strong outflow from the inner region of the accretion disks
\citep[][]{LaorDavis2014}, can also explain the apparently larger
light crossing time and flatter temperature profile.


\begin{figure}[htp]
\centering
\includegraphics[width=1\hsize]{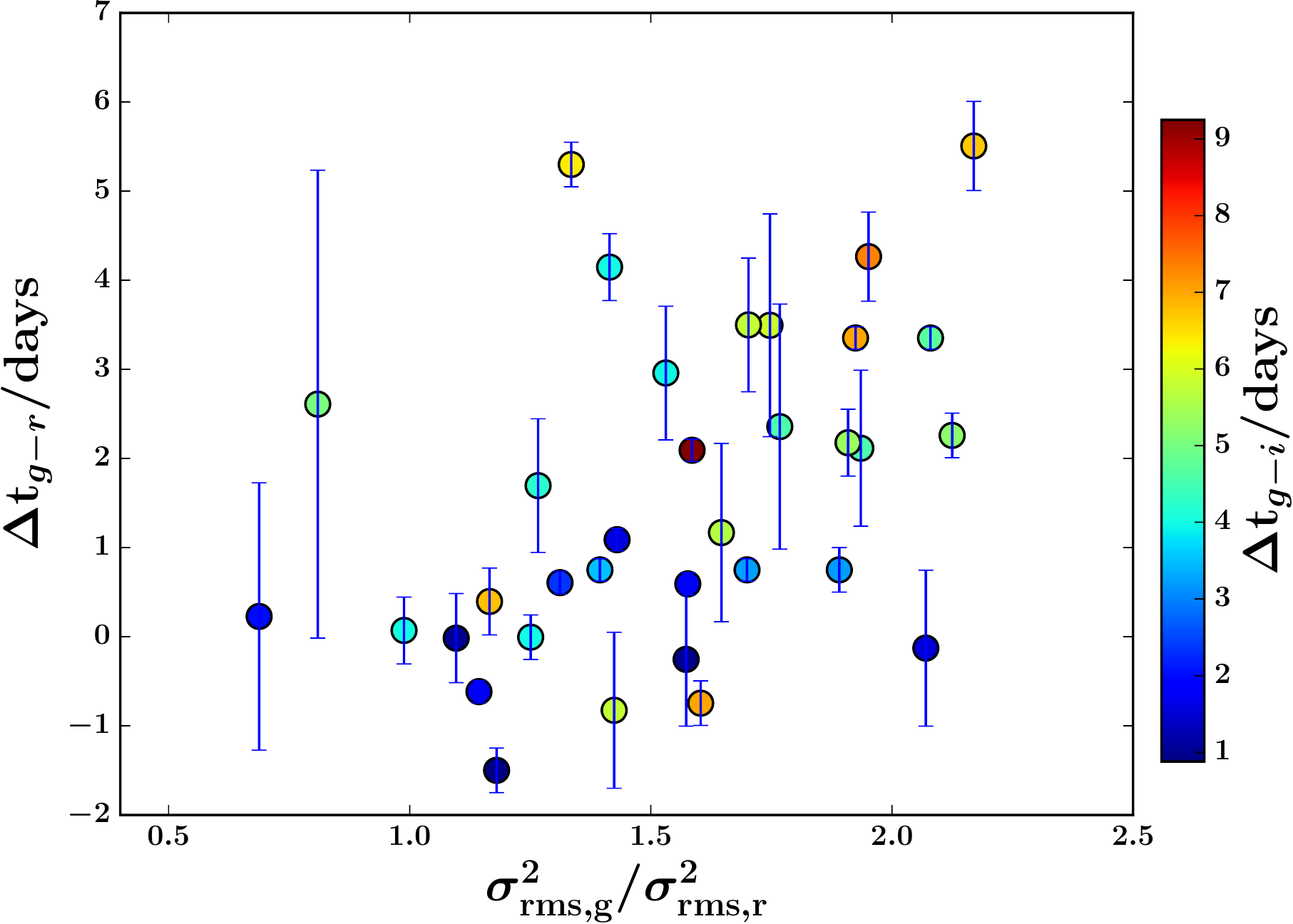}
\caption{
Correlation of $g-r$ lags ($\Delta t_{g-r}$) with the average ratio between 
the excess variance in the 
$g$ ($\sigma_{\rm rms,g}^2$) and $r$ ($\sigma_{\rm rms,r}^2$) 
bands for the $39$ \emph{cLD} quasars. Each data point 
is color coded by the $g-i$ lags ($\Delta t_{g-i}$). }
\label{negativelagsigma}
\end{figure}

\begin{figure}[htp]
\centering
\includegraphics[width=1\hsize]{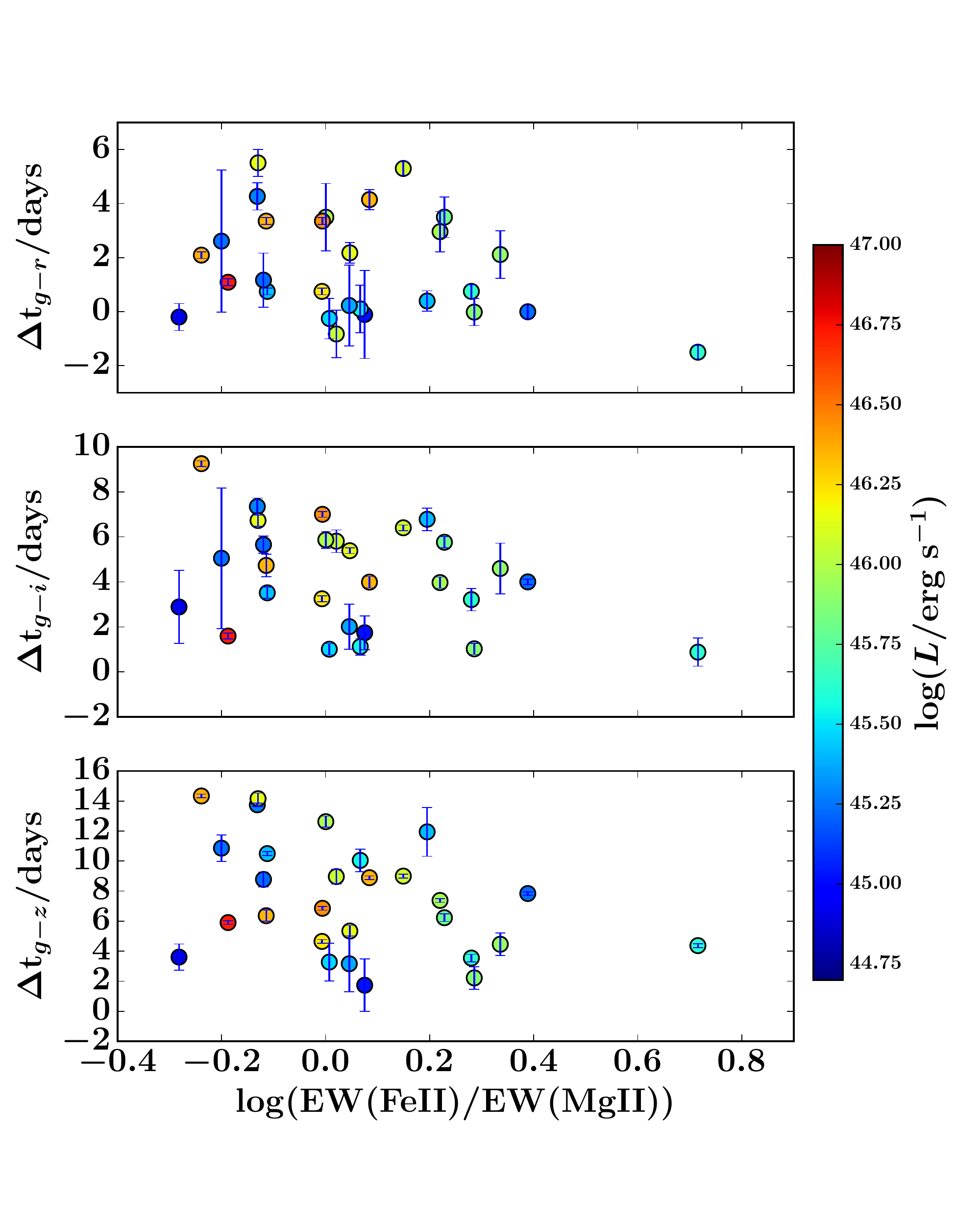}
\caption{Correlation of the $g-r$ (top panel, $\Delta t_{g-r}$), $g-i$ (middle panel, 
$\Delta t_{g-i}$) and $g-z$ (bottom panel, $\Delta t_{g-z}$) lags with the ratio between 
the equivalent widths of ultraviolet \ion{Fe}{2} and  \ion{Mg}{2} lines. 
Each data point is color coded with the bolometric luminosity.}
\label{laglineratio}
\end{figure}

\section{Discussions}
\label{sec:discussion}

\subsection{Comparison with the lags in NGC5548}

It is interesting to compare our results with the well-studied
NGC5548 \citep[][]{Edelsonetal2015,Fausnaughetal2016}. As we only
have four bands, we cannot constrain the radial temperature profile of
the disks over a wide range of wavelengths for each quasar, as in
NGC5548.  Instead, we can compare the detected and theoretically
expected lags in similar wavelength ranges. NGC5548 is at redshift
$0.017$ with the best estimated black hole mass $\sim 5.2\times
10^7\msun$ \citep[][]{Fausnaughetal2016}.  If we match the rest frame
wavelength at $z=1$ for the majority of our sample with NGC5548, the
Pan-STARRS $g$, $r$, $i$, $z$ bands corresponds to a wavelength
range of 2375\AA\ to 4350\AA\ for NGC5548. The reported lag between UVM2
(wavelength 2246\AA ) and B band (wavelength 4392\AA) for NGC5548 is
0.88 day \citep[][]{Edelsonetal2015}, while the theoretically expected
lag according to equation \ref{eqn:lags} is 0.34 day, if we assume the
accretion rate is $10\%$ of the Eddington accretion rate. This is consistent
with what we find for our sample in that the detected lag is larger
than the theoretically expected lag by a factor of 2.6.

\subsection{Lags with Unexpected Order}
\label{sec:negativelag}

A majority of the quasars show the short wavelength band leading the
long wavelength bands with positive $g-r$, $g-i$, and $g-z$ lags. 
On average, the lags are consistent with variabilities propagating outward via irradiation 
of the outer disk by the highly variable central radiation. 
Peaks in the
probability distributions for the stacked lags for our sample are 
always positive as shown in Figure \ref{LagAllSample}.

However, even in the subsample \emph{cLD} where well ordered lags with
a single dominant peak are found, there are still four quasars with
significantly negative values of $g-r$ lags as shown in the top panel
of Figure \ref{lagSubLumBH}, while the corresponding $g-i$ and $g-z$
lags are positive. We have also checked that when we measure the lags
between $r$ and $i$ bands for these quasars, we find positive
lags. Because quasars in each field have the same cadence, it will be hard to 
understand why only the four quasars in subsample \emph{cLD} show significant negative lags if they 
are artifacts of the cadence or the failure of {\sf JAVELIN}. 

For the 63 quasars in subsample \emph{iLD} having single dominant peaks 
but lags do not increase monotonically from $g-r$, $g-i$ to $g-z$, 14 of them 
have at least one significant negative lag, while 21 quasars 
have at least one band that does not follow this order significantly (at $2.35\sigma$ level). 

To test the statistical significance of the number of quasars with unexpected, we 
randomly draw lag signals from 200 quasars by assuming gaussian distributions 
with mean lags and standard deviations given by the stacked signals shown in 
Figure \ref{LagAllSample}. At $2.35\sigma$ level, there are no quasars with negative lags and 
there are only 7 quasars with lags that do not increase monotonically. All these numbers 
are much smaller than what we get from the data. 

If these lags are physical signals, 
this suggests that there could still be drivers of the light-curve variability 
for these quasars located at the outer part of the disk
where longer wavelength radiation is emitted.
If the variability is caused by a driver at a particular radius and
propagates outwards, we expect the variation amplitude to decrease as
the perturbation travels. In order to test the location of the driver,
we calculate the normalized excess variance in the $g$ 
($\sigma_{\rm rms,g}^2$) and $r$ bands ($\sigma_{\rm rms,r}^2$) in each season 
and calculate the mean ratio $\sigma_{\rm rms,g}^2/\sigma_{\rm rms,r}^2$ 
of the four observational seasons for each quasar in subsample \emph{cLD},
which is plotted against the $g-r$ lags in Figure
\ref{negativelagsigma}. There is a strong correlation between $\Delta
t_{g-r}$ and $\sigma_{\rm rms,g}^2/\sigma_{\rm rms,r}^2$ with a
Pearson correlation coefficient $0.43$ with corresponding $p$ value $0.01$
 and a Spearman correlation
coefficient $0.47$ with corresponding $p$ value $0.005$. 
For the quasars with positive $\Delta t_{g-r}$,
the fact that $\sigma_{\rm rms,g}^2/\sigma_{\rm rms,r}^2$ is larger as
$\Delta t_{g-r}$ increases is consistent with the picture that the
disturbance propagates from $g$ band to $r$ band and gets weaker as it
travels.  The further it propagates, the more the signal is damped. For
the quasars with significant negative $\Delta t_{g-r}$, 
$\sigma_{\rm rms,g}^2/\sigma_{\rm rms,r}^2$ is smaller and gets close to
$1$. Particularly in the first season, we find that the ratio $\sigma_{\rm
  rms,g}^2/\sigma_{\rm rms,r}^2$ is smaller than $1$ for these
quasars. This is consistent with the suggestion that the driver is
closer to the region where $r$ band radiation is emitted for the
quasars with negative $g-r$ lags.

It is hard to understand the existence of these negative lags around
the light crossing time scales in the context of the standard thin disk
model, which predicts that the scale height of the radiation pressure 
dominated inner region of the black hole accretion disk is a constant
for different radii \citep[][]{ShakuraSunyaev1973}. The X-ray and far
UV emission in quasars, which are thought to drive the variability,
are believed to be produced by the corona on top of the accretion
disk \citep[e.g.,][]{HaardtMaraschi1991,HaardtMaraschi1993}, and are also
found to be located in a compact region near the innermost stable
circular orbit \citep[][]{Chartasetal2009,ReisMiller2013}. In order
for an off-center region to produce significant variability, a special
mechanism is required to operate at the $r-$band region and change the
disk thickness there so that it can irradiate nearby radii.  Recently,
\cite{Jiangetal2016} pointed out that the iron opacity bump, which
would exist around $1.8\times 10^5$ K inside the disk, plays an
important role to stabilize the accretion disks in AGNs and change the
disk thickness.  This is a promising mechanism as $r-$band radiation
is expected to be emitted from a region around $26$ Schwarzschild
radii in our sample according to the estimate in Section
\ref{sec:theory}, which is also the location where the iron opacity bump
is expected. This mechanism also predicts significant outflows launched
from this region, which can be
tested with future observations.

Although only $16\%$ of quasars in our sample have significant, well
ordered lags for all $g$, $r$, $i$, $z$ bands individually, the
stacked lags of the whole sample are all positive and increase with
increasing wavelength differences. This is true even if we do not
include subsample \emph{cLD} during the stacking process. It suggests
that this variability component, which is consistent with the picture
of irradiation by the central source, exists for most
quasars. However, for each individual quasar, this may not be the
dominant component around the light crossing time scales. For example,
strong off-center disturbances caused by localized regions around the
iron opacity peak \citep{Jiangetal2016} can be one candidate to cause
the variability. Locations of the drivers in this case will change
more significantly with black hole mass and accretion rate, which may
explain why these signals do not show up in the stacking process.

\subsection{The Effects of Metallicity}

It has been suggested that the line ratio \ion{Fe}{2}/\ion{Mg}{2} is a
good proxy of metallicity in the broad line gas of quasars, and
probably the accretion disk, although the uncertainty is large
\citep[][]{Kurketal2007, DeRosaetal2011}.  As the iron opacity bump
will increase with higher metallicity, the modifications in disk
structure compared with the standard thin disk model 
\citep{Jiangetal2016} will be more significant, as long as the metallicity 
is larger than the solar value. Therefore the
inter-band lags may also show some correlation with the ratio between
the equivalent widths of ultraviolet \ion{Fe}{2} and \ion{Mg}{2},
which is shown in Figure \ref{laglineratio} for the subsample
\emph{cLD}. Despite large scatter, all three lags $\Delta t_{g-r}$,
$\Delta t_{g-i}$, and $\Delta t_{g-z}$ show weak anti-correlations with
EW(\ion{Fe}{2})/EW(\ion{Mg}{2}), particularly when
EW(\ion{Fe}{2})/EW(\ion{Mg}{2}) is larger than 1.  The Pearson
correlation coefficients between the equivalent width ratio and
$\Delta t_{g-r}$, $\Delta t_{g-i}$, $\Delta t_{g-z}$ are $-0.34$,
$-0.33$, $-0.30$, while the corresponding Spearman correlation
coefficients are $-0.20$, $-0.28$ and $-0.35$. The $p$ values 
for Pearson correlation coefficients are $0.07, 0.08$ and $0.11$.   Each data point is
color coded by luminosity in Figure \ref{laglineratio}, which
shows no clear trend between luminosity and the equivalent ratio. This
suggests that the anti-correlation between lag and
EW(\ion{Fe}{2})/EW(\ion{Mg}{2}) is independent of the weak correlation
between lag and luminosity shown in Figure \ref{lagSubLumBH}.  If the
equivalent width ratio is a good indicator of metallicity, this indeed
suggests that at large enough metallicity, the structure of
the disk may change and affect the inter-band lags.  However, more theoretical
studies are clearly needed to understand each type of lag behavior as
described in Section \ref{sec:wholesample}.

\subsection{Uncertainties in the Thin Disk Model}

There are still some uncertainties in the simple disk model we compare
to, which may affect the predicted lags in equation
\ref{eqn:lags}. The parameter $f_i$ represents the heating of the disk
by the central radiation. We set it to be 1 for
simplicity\footnote{The parameter $f_i$ is equivalent to $1+\kappa/3$
  in \cite{Fausnaughetal2016}, which was chosen to be $4/3$.}, which
means the disk temperature is still determined by the local viscous
heating. Because the lags are only proportional to $f_{i}^{1/3}$, in
order to make the predicted lags three times larger, $f_i$ needs to be
increased to $27$, which means the disk temperature is completely
determined by the irradiation. However, this is unlikely, particularly
in the outer part of the disk because the central radiation flux will
drop with distance $R$ as $R^{-3}$ in the lamp post model with a 
thin disk geometry. There is also no reason to believe
that the disk temperature will have the same radial profile as in the
standard thin disk model, which is what we assume in equation
\ref{eqn:lags}.  The predicted lags also have a weak dependence on the
black hole mass. If the discrepancy is caused by errors in the
estimated black hole mass, the masses of all the quasars need to be
systematically underestimated by a factor of 27 in order to increase
the predicted lags by a factor of 3. Although the single-epoch virial
method we use to estimate the black hole mass is very uncertain, the
systematic error is unlikely to be so large \citep[][]{Shen2013}. If
the black hole mass increased by a factor of 27, there would be many
black holes with mass larger than $10^{10}\msun$ and the
Eddington ratios for most of the quasars would be smaller than $\sim
1\%$, neither of which is likely. The inclination of the disk will also
change the line of sight distance at different locations of the
disk. On the near side of the disk, the outer part of the disk is
closer to us while on the far side of the disk, the inner part is
closer to us. This will just broaden the lag signals and the 
mean lag values will be unaffected \citep[][]{Starkeyetal2016}.

\subsection{Extension to Different Redshifts}

We selected particular redshift ranges to avoid significant
contamination by broad emission lines in the Pan-STARRS filters.  This
also limits the size of our sample and the radial range of the
accretion disks we can probe. In principle, when we fit the
light-curves based on DRW models, we can use more than one component
in each band to fit both the continuum and lines simultaneously. A
similar technique has been demonstrated by \cite{CheloucheZucker2013} (see also \citealt{Zuetal2016})
to separate the continuum-continuum and continuum-line lags. In this
way, lags of quasars in different redshift ranges can be studied.  We
will apply this technique to the Pan-STARRS data in the near future.

\section{Summary}

\begin{figure}[htp]
\centering
\includegraphics[width=1\hsize]{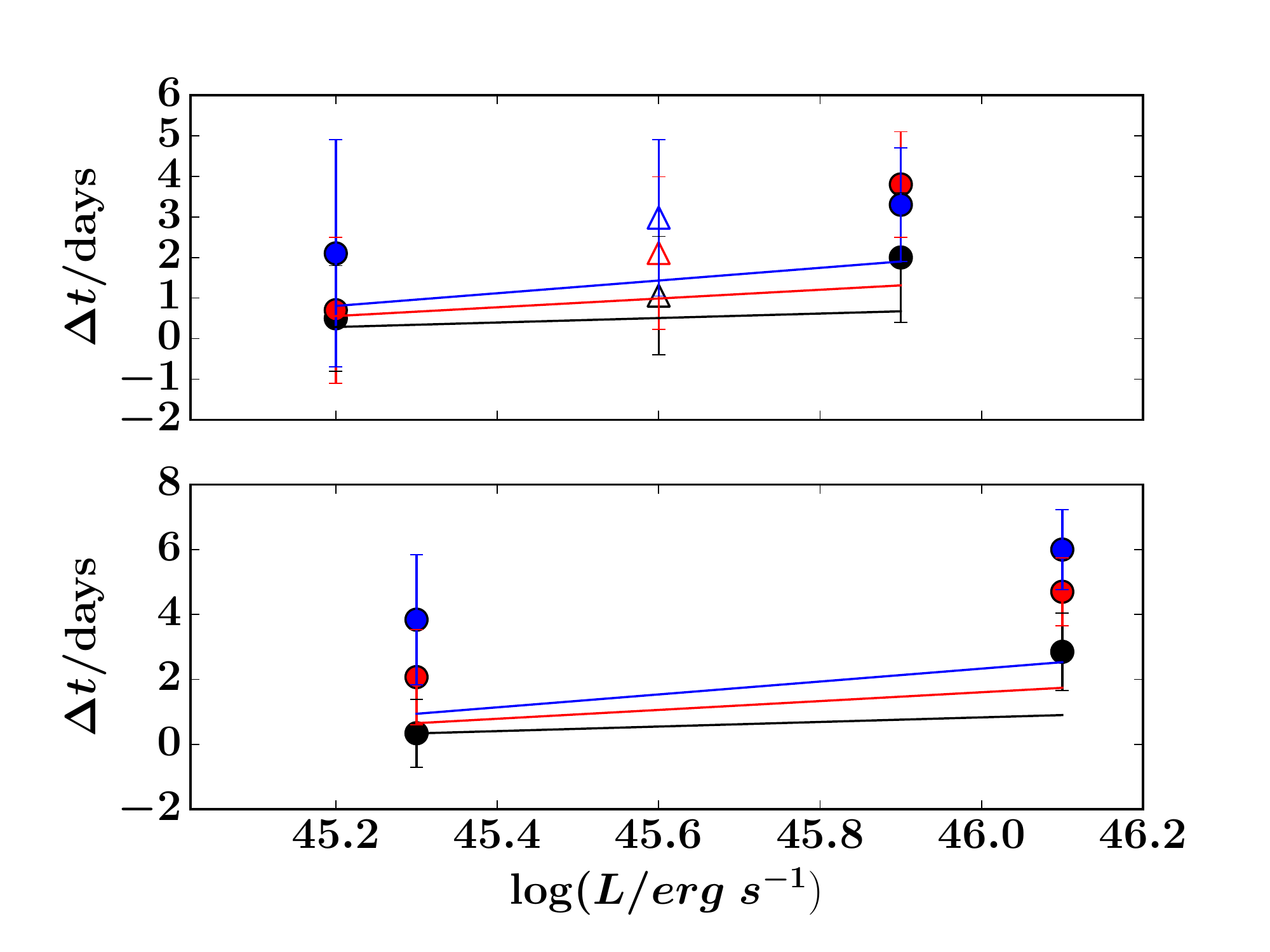}
\caption{
\emph{Top:} summary of the stacked lag signals with median luminosity 
for the whole sample. The open
black, red and blue triangles are the stacked $g-r$, $g-i$ and $g-z$
lags as shown in Figure \ref{LagAllSample}.  The
filled black, red, and blue circles are the stacked $g-r$, $g-i$ and
$g-z$ lags for two luminosity bins with the
theoretically expected values for each luminosity bin connected by the
black, red and blue lines. \emph{Bottom}: 
the corresponding stacked lags of the
subsample \emph{cLD} for two luminosity bins as shown in Figure
\ref{lagSubLum}.}
\label{summaryplot}
\end{figure}

In summary, we have used more than four years of light curves from the
Pan-STARRS medium deep fields to detect continuum-band lags in 200
quasars.  The mean lags between the $g-$band and the $r$, $i$, and $z$
bands for the whole sample are $1.1$, $2.1$ and $3.0$ days.  There are
$39$ quasars showing significantly detected lags that increase
toward redder bands, as expected if the lags
correspond to the light crossing times across different radii of the
accretion disks when the outer part of the disk is irradiated by the
central source. The detected lags are systematically larger than
the expected values based on the standard thin disk models by a factor
of $2-3$, which cannot be explained by uncertainties in the
measurements (for example black hole mass) or the thin disk model (for
example the inclination).  The stacked lags and theoretically expected
values are summarized in Figure \ref{summaryplot}.  This is consistent
with the recent results for NGC5548 and microlensing measurements. The
correlations between the $g-r$, $g-i$ and $g-z$ lags are also
significantly different from thin disk model predictions, particularly
for quasars with larger lags and higher luminosities.

The detected lags are found to increase with increasing luminosity,
which is also clearly shown in Figure \ref{summaryplot}.  This is
probably because accretion disk sizes are larger for high luminosity
quasars. We also find evidence that the lags decrease with increasing
ratio EW(\ion{Fe}{2})/EW(\ion{Mg}{2}), particularly when this ratio is
large. This may indicate that the accretion disk structure is
changed in quasars with higher metallicity, probably because of the
effects of the iron opacity bump \citep[][]{Jiangetal2016}. There are also four quasars in 
subsample \emph{cLD} with
significant negative lags between the $g$ and $r$ bands and we find the
correlation that the ratios between the excess variance in $g$ and
$r$ bands generally increase with increasing $g-r$ lags. This
indicates that some quasars may have strong off-center variability
that will complicate the lag signals.

It will be interesting to carry out the same experiment with more  data 
at different redshifts, which will allow us to probe a larger radial range of the accretion disk. 
For the quasars with lags that are consistent with the lamp post model, the 
Continuum Reprocessed AGN Markov Chain Monte Carlo (CREAM) \citep[][]{Starkeyetal2016} 
model can be used to constrain the properties of the accretion disks (such as inclination and 
mass accretion rate). 
The correlations we find between the lags and physical properties of the accretion disks 
will be significantly improved with better sampled data and more quasars. This 
will be one interesting application of LSST data. Better data with regular $\sim$ one day cadence 
will also be able to tell whether the lags with unexpected orders are physical or not.

\section*{Acknowledgements}
Y.F.J. and A.P. are supported by NASA through
Einstein Postdoctoral Fellowship grant number PF3-140109 and 
PF5-160141 awarded by the Chandra X-ray Center. 
YS acknowledges support from an Alfred P. Sloan Research Fellowship 
and WNB acknowledges support from NSF grant AST-1516784. 
This research is also supported in part by the National Science Foundation 
under NSF PHY-1125915. 

The Pan-STARRS1 Surveys (PS1) have been made possible through
contributions by the Institute for Astronomy, the University of
Hawaii, the Pan-STARRS Project Office, the Max-Planck Society and its
participating institutes, the Max Planck Institute for Astronomy,
Heidelberg and the Max Planck Institute for Extraterrestrial Physics,
Garching, The Johns Hopkins University, Durham University, the
University of Edinburgh, the Queen's University Belfast, the
Harvard-Smithsonian Center for Astrophysics, the Las Cumbres
Observatory Global Telescope Network Incorporated, the National
Central University of Taiwan, the Space Telescope Science Institute,
and the National Aeronautics and Space Administration under Grant
No. NNX08AR22G issued through the Planetary Science Division of the
NASA Science Mission Directorate, the National Science Foundation
Grant No. AST-1238877, the University of Maryland, Eotvos Lorand
University (ELTE), and the Los Alamos National Laboratory.

Funding for SDSS-III has been provided by the Alfred P. Sloan Foundation, 
the Participating Institutions, the National Science Foundation, and the U.S. Department of Energy Office of Science. 
The SDSS-III web site is http://www.sdss3.org/.

SDSS-III is managed by the Astrophysical Research Consortium for the Participating Institutions of the SDSS-III 
Collaboration including the University of Arizona, the Brazilian Participation Group, Brookhaven National Laboratory, 
Carnegie Mellon University, University of Florida, the French Participation Group, the German Participation Group, 
Harvard University, the Instituto de Astrofisica de Canarias, the Michigan State/Notre Dame/JINA Participation Group, 
Johns Hopkins University, Lawrence Berkeley National Laboratory, Max Planck Institute for Astrophysics, 
Max Planck Institute for Extraterrestrial Physics, New Mexico State University, New York University, 
Ohio State University, Pennsylvania State University, University of Portsmouth, Princeton University, 
the Spanish Participation Group, University of Tokyo, University of Utah, Vanderbilt University, University of Virginia, 
University of Washington, and Yale University.

\bibliography{QuasarLag}

\end{CJK*}

\end{document}